\documentclass[modern, linenumber]{aastex701}

\usepackage{siunitx}
\usepackage{booktabs}%
\usepackage{graphicx}%
\usepackage{multirow}%
\usepackage{amsmath,amssymb,amsfonts}%
\usepackage{amsthm}%
\usepackage{mathrsfs}%
\usepackage{colortbl}
\usepackage{textcomp}%
\usepackage{algorithmicx}%
\usepackage{algpseudocode}%
\usepackage{listings}%
\usepackage[version=4]{mhchem}
\usepackage{threeparttablex} % for footnote in table
\usepackage{array}
\newtheorem{assumption}{Assumption}

\begin{document}
\title{The Physical Nature of Regolith on Icy Moons}
%Catchy title
%What does the uppermost surface of icy moons look like?
%Alternative more serious but less catchy title:
%\title{The Low Thermal Inertia of Icy Moons: Implications on Surface Porosity, Grain Size, and Regolith Structure}

%\correspondingauthor{Cyril Mergny}

\author[0009-0002-1910-6991]{Cyril Mergny}
\affiliation{European Space Agency (ESA), European Space Astronomy Centre (ESAC), E-28692 Villanueva de la Ca\~nada, Madrid, Spain}
\email{cyril.mergny@esa.int}

\author[0000-0001-5971-0056]{Thomas Cornet}
\affiliation{European Space Agency (ESA), European Space Astronomy Centre (ESAC), E-28692 Villanueva de la Ca\~nada, Madrid, Spain}
\email{thomas.cornet@esa.int}

\author[]{Alice Le Gall}
\affiliation{LATMOS/IPSL, Sorbonne Université, UVSQ, CNRS, Paris, France}
\email{}

\author[]{Guillaume Cruz-Mermy}
\affiliation{European Space Agency (ESA), European Space Astronomy Centre (ESAC), E-28692 Villanueva de la Ca\~nada, Madrid, Spain}
\email{}

\author[]{Lucas Lange}
\affiliation{NASA Postdoctoral Program Fellow at Jet Propulsion Laboratory (JPL), California Institute of Technology, Pasadena, CA, USA.}
\email{lucas.lange@jpl.nasa.gov}

\author[]{Tina Rückriemen-Bez}
\affiliation{University of Münster, Institut für Planetologie, Münster, Germany.}
\email{}

\author[]{Bastian Gundlach}
\affiliation{University of Münster, Institut für Planetologie, Münster, Germany.}
\email{}

\author[]{Paula Heitmann}
\affiliation{University of Münster, Institut für Planetologie, Münster, Germany.}
\email{paula.heitmann@uni-muenster.de}

\author[0009-0007-6508-8719]{Moritz Goldmann}
\affiliation{University of Münster, Institut für Planetologie, Münster, Germany.}
\email{moritz.goldmann@uni-muenster.de}

\author[]{Paul Hayne}
\affiliation{Laboratory for Atmospheric \& Space Physics, University of Colorado Boulder}
\affiliation{Department of Astrophysical \& Planetary Sciences, University of Colorado Boulder}
\email{paul.hayne@colorado.edu}

\author[0000-0002-1655-0715]{Apurva V. Oza}
\affiliation{Division of Geological and Planetary Sciences, California Institute of Technology, Pasadena, CA, USA}
\email{oza@caltech.edu}

\begin{abstract}

Estimating surface properties such as porosity and grain sizes is key for planning lander missions and landing site selection on icy moons.
However, spaceborne instruments do not measure the regolith properties directly: instead, they record proxy measurements such as thermal flux which are then interpreted through modeling to estimate thermal inertia, porosity, grain size, etc...

A striking conclusion from all thermal measurements that probed the \textit{uppermost} surface (first millimeters) of icy moons is they all show an exceptionally low thermal inertia, ranging from $\qtyrange[range-units=single]{9}{20}{J.m^{-2}.K^{-1}.s^{-1/2}}$.
This value is orders of magnitude lower than that of bulk hexagonal water ice, $\sim \SI{2000}{J.m^{-2}.K^{-1}.s^{-1/2}}$ at these temperatures.
We demonstrate that a regolith thermally dominated by hexagonal water ice may only achieve  such thermal inertia through a combination of extremely high porosity ($>80\%$), small grain radii ($<\SI{1}{mm}$) and an unconsolidated regolith (minimal contact area between grains), consistent with previous photometry and spectroscopy studies.

For the Galilean moons, deeper thermal observations ($> \SI{1}{cm}$) have revealed higher thermal inertia $ > ~\SI{50}{J.m^{-2}.K^{-1}.s^{-1/2}}$, indicating that the regolith compacts over centimeter scales.
Since gravity has no effect on compaction on such scale, we propose three formation scenarios to account for vertical layering: deposition cover, degradation by impactors, and temperature gradient metamorphism. 

We discuss how monodisperse grains can reach such extreme porosities and provide examples of experimental analogs that could best represent the regolith.
We propose that high porosity regolith are favored on icy moons due to the adhesive nature of water ice and their low-gravity environment.

%In turn, defining the range of allowed porosities and grain sizes will improve linear unmixing and Bayesian methods, by providing strong constraints on the parameter space to explore. 
%Tighter or looser constraints can be derived depending on the assumptions underlying the various models found in the literature, which are also discussed. 
%including for future MAJIS (JUICE) and MISE (Europa Clipper) spectrometers, as well as for the planning of lander operations and/or landing site selection on such highly porous surfaces (e.g., Voyage2050, ESA’s L4 mission).
%While a lower thermal inertia could be attributed to insulating materials (\textit{e.g.}, dust or amorphous ice phases), regions of pure crystalline ice also exist on these bodies and yet they still present such low thermal inertia near the surface.

\end{abstract}

%\begin{keywords}
%Icy Regolith, Remote Sensing, Surface Interpretation
%\end{keywords}

% ------ Main Section -------------------------------------
\section{Introduction}

From Earth-telescopes in the 1970s \citep{Hansen1973, Morrison1973} to recent ALMA observations \citep{Trumbo2018, Kleer2021, Camarca2023, Thelen2024} and spacecraft data from Voyager's Visual and Infrared Imaging Spectrometer  (VIRS) \citep{Flasar2004}, Galileo's Photopolarimeter-Radiometer (PPR) \citep{Russell1992}, and Cassini's Composite Infrared Spectrometer (CIRS) \citep{Hanel1977}, the emitted thermal radiation from the surfaces of all icy moons in the solar system has been collected \citep{Ferrari2018}.
Since these measurements became available, using both analytical and numerical approaches, models have sought to derive the thermal properties of these surfaces and in particular their thermal inertia (denoted as $\Gamma$, in $\SI{}{J.m^{-2}.K^{-1}.s^{-1/2}}$, with units omitted hereafter for readability).
Thermal inertia is defined from the fundamental heat properties: conductivity, density and heat capacity, as:
\begin{equation}
    \Gamma = \sqrt{k\rho  c_p}
    \label{eq:inertia}
\end{equation}
where $k$ is the thermal conductivity of the considered medium, $\rho$ its density and $c_p$ the specific heat capacity of the material. 

The inverse methods wich have been used to deduce surface properties are specific to the depths probed by each instruments. 
If the regolith is vertically heterogeneous, its properties, like thermal inertia will vary with depth;  this is often referred to as \textit{surface layering}.
In such situation, the same surface, observed under identical conditions, may show very different thermal inertia values depending on the probing depth.
For infrared instrument, the probing depth is considered to be the depth affected by thermal variations, called the thermal skin depth and defined as
\begin{equation}
    \delta_{\mathrm{th}}(P, \phi) = \sqrt{\dfrac{k}{\rho c_{\mathrm{p}} } \dfrac{P}{\pi}}=\frac{\Gamma}{\rho c_{\mathrm{p}}}\sqrt{\dfrac{P}{\pi}}
\end{equation}
where $P$ is the period of the thermal cycle. 
As noted by \citet{Ferrari2018}, planetary surfaces can experience three main types of thermal cycles: eclipses, diurnal cycles, and seasonal cycles, with durations differing by orders of magnitude, ranging from hours to days or months and years. 
For infrared observations of icy moons, this typically results in probing depths of millimeters to centimeters for eclipses and diurnal cycles, and meters for seasonal cycles.
For observations at longer wavelengths, such as microwave, the thermal emission may originate from depths of centimeters up to meters. \citep{Ferrari2018}.

A striking commonality emerges from all thermal inversions of icy moons (see Table \ref{tab:lowTI}): the topmost surface (\textit{i.e.} the first millimeters depth) consistently show exceptionally \textit{Low Thermal inertia across all Icy Moons}, hereafter referred as $\Gamma_{\mathrm{LTIM}}$ (a summary table of all parameters and notations used in this paper can be found in Table~\ref{tab:sumtab}).
For the Jovian icy moons, the thermal inertia in the first millimeters depth is always $\Gamma_{\mathrm{LTIM}}< \SI{15}{}$. For Saturnian moons, $\Gamma_{\mathrm{LTIM}} \leq \SI{20}{}$ but can reach even values as low as $\Gamma_{\mathrm{LTIM}}< \SI{10}{}$ up to centimeters deep.
These values are orders of magnitude lower than those of bulk  (\textit{i.e.} solid / non-porous) hexagonal water ice, hereafter named $I_{\mathrm{h}}$, which, based on measured thermal properties, should approach \(\Gamma_{\mathrm{b}}(I_{\mathrm{h}}) = \SI{2000}{}\) (see derivation in Section \ref{sec:bulk}).
This raises the question: what kind of structure require these icy surfaces to have a thermal inertia two orders of magnitude lower than bulk water ice ?
%we take 15MKS falls withing error bars of all, and lower value will give even tighter restrains.
%smaller than 30 MKS were derived from Cassini CIRS observations over the SPT (Howett et al., 2010)

\begin{table}[htbp]
    \centering
    \begin{tabular}{l l l l l l}
        \toprule
        {Body} & Thermal inertia  & {Density depth}  & Type$^{*}$ & Instrument & References  \\
            &  ($\SI{}{J.m^{-2}.K^{-1}.s^{-0.5}}$) & ($\SI{}{\kilo\gram.m^{-2}}$) & &  \\
        \midrule
        %\textit{Jovian icy moons} \\
        Europa & $14 \pm 5$ & $\SI{0.9}{}$  & Eclipse & Earth-T$^{**}$  & \citenum{Hansen1973} \\
               & $57 \pm 17$ & $\SI{21.2}{}$ & Diurnal & PPR & \citenum{Lange2026} \\ 
               & $70$ & $\SI{26.1}{}$ & Diurnal & PPR & \citenum{Spencer1999} \\ 
               & $45-150$ & $\SI{26.1}{}$ & Diurnal & PPR & \citenum{Rathbun2010} \\ [4pt]
               & $95$ & & Microwave & ALMA$_6$ & \citenum{Trumbo2018} \\ [4pt]
               %Trumbo says range 40 to 300
               & $56-184$ & & Microwave & ALMA$_3$ & \citenum{Thelen2024} \\ [4pt]
               %Thelen says betwen 8 and 20 cm
        Ganymede & $12 \pm 3$ & $ 0.8 $ & Eclipse & Earth-T$^{**}$ & \citenum{Hansen1973} \\
               & $14 \pm 3$ & $ 0.9 $ & Eclipse & Earth-T$^{**}$  &\citenum{Morrison1973} \\
               & $70 \pm 20$ & $ 37.0 $ & Diurnal & PPR & \citenum{Spencer1987} \\[4pt]
               & $400-800$ & & Microwave & ALMA$_{3, 6, 7}$ & \citenum{Kleer2021} \\ [4pt]
        Callisto & $10 \pm 1$ & $ 0.7 $  & Eclipse & Earth-T$^{**}$  & \citenum{Morrison1973} \\
               & $50 \pm 10$ & $ 40.4 $ & Diurnal & PPR & \citenum{Spencer1987} \\[4pt]
                & $15-2000$ & & Microwave & ALMA$_{3, 6, 7}$ & \citenum{Camarca2023} \\ [4pt]
        %\textit{Saturnian moons} \\
        \midrule
        Mimas & $19^{+57}_{-9}$ & $ 3.6 $ & Diurnal & CIRS & \citenum{Howett2010} \\[4pt]
        Enceladus & $15^{+24}_{-9}$ & $3.5$  & Diurnal &  CIRS & \citenum{Howett2010}\\[4pt]
        Tethys & $9^{+10}_{-4}$ & $2.4$ & Diurnal &  CIRS & \citenum{Howett2010} \\[4pt]
        Dione & $9-16$ & $0.3-0.6$ & Eclipse &  CIRS & \citenum{Howett2022}\\
        & $11^{+18}_{-6}$ & $3.6$ & Diurnal &  CIRS & \citenum{Howett2010} \\[4pt]
        Rhea & $8_{-5}^{+12^{\mathrm{l}}}/9^{+9^{\mathrm{t}}}_{-5}$ & $3.6$ & Diurnal & CIRS & \citenum{Howett2010} \\
         & $1-46$ & $0.4-19.3$ & Seasonal & CIRS & \citenum{Howett2016} \\
        & $>60$ &  & Microwave & RADAR/Cassini & \citenum{Bonnefoy2020}\\[4pt]
        Iapetus & $20^{+13^{\mathrm{l}}}_{-8}/14^{+7^{\mathrm{t}}}_{-8}$  & $29.9$ & Diurnal & CIRS & \citenum{Howett2010} \\
         & $11-25$  & $19.3-43.9$ & Diurnal & CIRS & \citenum{RiveraValentin2011} \\
        & $>200$  &  & Microwave & RADAR/Cassini & \citenum{LeGall2014} \\[4pt]
        Phoebe & $20-25$ & $ 2.7$ & Diurnal & CIRS & \citenum{Howett2010} \\[4pt]
        \bottomrule
    \end{tabular}

    \begin{threeparttable}
    \begin{tablenotes}
    \footnotesize
        \item[*] Type of observations: during eclipse, during a diurnal cycle or a seasonal cycle.
        \item[**] Earth-Telescopes: Hale observatory
telescopes \citep{Hansen1973} and  Mauna Kea Observatory telescope \citep{Morrison1973}.
        \item[l, t] For Rhea and Iapetus, values are given both for the leading and trailing hemispheres.
    \end{tablenotes}
    \end{threeparttable}

    \caption{Thermal inertia values estimated on icy moons, modified from \cite{Howett2010} and \citet{Ferrari2018}.
    Following \citet{Cooper2001} we express depths in $\SI{}{kg.m^{-2}}$, this simplifies discussion of the density-dependence in the porous regolith. To retrieve the thermal skin depth, one can divide by the density $\rho(\phi)$ of the regolith.
    }
    \label{tab:lowTI}
\end{table}

%While skin depth can be expressed independently of the density (in $\SI{}{kg.m^{-2}}$), here they are expressed by assuming a porosity of $90\%$, which gives a more accurate representation of the probed depth, given the results of our study.

A commonly accepted hypothesis is that the surface is made of an icy regolith with important porosity. Such porosity would reduce the conductivity and density of the ice, thereby lowering its thermal inertia \citep{Lellouch2013}.
However, \citep{Ferrari2016} argued that porosity alone may not be sufficient to explain these low thermal inertia values and that other insulating materials must be invoked. Spectroscopy has revealed that icy moons' surfaces contain other compounds than crystalline water ice. These other elements could lower the regolith's thermal conductivity. 
A promising candidate to lower thermal inertia is amorphous water ice, hereafter named $I_{\mathrm{a}}$. Based on this hypothesis, \cite{Ferrari2016} showed that the bulk thermal inertia of amorphous water ice, in the $20$ to $\SI{120}{K}$ temperature range, varies in the order of $\Gamma_{\mathrm{b}}(I_{\mathrm{a}}) = 30-\SI{500}{}$. This is about an order of magnitude lower than its crystalline counterpart $\Gamma_{\mathrm{b}}(I_{\mathrm{h}}) \sim 2000$ which does not vary significantly with temperature. This allows amorphous ice to match $\Gamma_{\mathrm{LTIM}}$ without requiring an extremely high porosity.

Without invalidating the results of \textcite{Ferrari2016}, we argue that amorphous ice, due to its metastable nature, cannot universally explain the low thermal inertia observed on icy moons.
Instead, our study takes a different direction, by exploring what type of regolith would require hexagonal water ice to be compatible with such low thermal inertia values.

Indeed, while spectroscopic measurements have detected traces of amorphous water ice on some of these moons, crystalline water ice appears to be the predominant form \citep{Mastrapa2013}.
On Europa, and even more so on Ganymede and Callisto, temperatures are high enough for ice to be predominantly in its crystalline phase \citep{Schmitt1989, Hansen2004, Mergny2025cry}. 
Although regions of Europa experience radiation fluxes of charged particles coming from Jupiter's magnetosphere that can be high enough for amorphization to dominate crystallization. Nevertheless, this amorphization process is less efficient at higher temperatures and spectroscopy and numerical modeling have shown that most of Europa's surface is made of crystalline ice \citep{Berdis2020, Mergny2025cry}.
%Also at depth deeper than 1mm becomes almost fully crystalline everywhere!}
On Ganymede, only the poles, more exposed to the Jupiter radiation, show signs of amorphous ice \citep{BockeleeMorvan2024}. On Callisto, based on typical water ice crystallization timescales \citep{Schmitt1989, Kouchi1994}, temperatures are too high for amorphous water ice to be stable over geological timescales.
For Saturnian satellites, even Enceladus, whose surface temperatures are usually much colder  \citep[around 70K,][]{Spencer2006, Howett2010} which allows ice to remain locked into its amorphous ice, spectroscopy has shown ice to be predominantly composed of crystalline water ice \citep{Clark2012, Filacchione2012}. 
%This can be explained if the ice is formed directly in its crystalline state, meaning that during the formation of ice grains in the plumes, temperatures are sufficiently high for ice to bypass the amorphous phase. 
While traces of amorphous water ice are present on some of these surfaces, most of these observed icy surfaces therefore seem to host water ice in its crystalline state.

Additionally, the $\Gamma_{\mathrm{LTIM}} \approx 15$ values in Table~\ref{tab:lowTI} represent a global average over the full hemisphere of a moon.
%I've verified for Hansel at least
However, colder areas contribute less to the observed thermal emission because the radiative gray-body emission from these surfaces scales with $\propto T^4$, when seen by infrared instruments.
Thus, hotter regions will dominate the emitted flux compared to colder regions of equal area where amorphous ice is more likely to be present.
As a result, regions where amorphous ice might be stable (i.e., colder ones) contribute less to the spectra that were used to infer the low thermal inertia reported in Table \ref{tab:lowTI}, making, in turn, amorphous water ice a less likely candidate to explain such low thermal inertia.

Finally, spectroscopy showed that the surfaces of these icy moons are not composed of pure water ice and contain contaminants such as sulfuric acid hydrate \citep{Carlson1999, Carlson2002}, hydrated sulfates \citep{McCord1998,McCord2002,Dalton2007}, chlorinated compounds \citep{Brown2013,Trumbo2019}, \ce{CO2} \citep{Brown2006,Combe2019, Villanueva2023, Trumbo2023}, \ce{NH3}, \citep{Grundy1999,Emery2005}, \ce{SO2}
and organic compounds \citep{McCord1997,McCord1998,McCord2001,BockeleeMorvan2024}. 
These contaminants will influence the thermal properties of the icy regolith, which could potentially explain these low thermal inertia values. However, it remains unclear whether these compounds act as insulating or conducting materials within the regolith, which we discuss in Section~\ref{sec:limits}.
Nonetheless, regions of nearly pure crystalline water ice do exist on these moons \citep{Hansen2004}, and if $\Gamma_{\mathrm{LTIM}}$ values are correct, they give strong constrains on the regolith thermal properties to be orders of magnitude lower than that of bulk water ice.

The main question addressed in this paper is therefore: assuming that thermal inertia values as low as those in Table~\ref{tab:lowTI} are correct (Assumption \ref{as:correctTI}) and that the surface is thermally governed by hexagonal water ice, what does this reveal about the structure of the regolith?

\begin{table}[htbp]
    \centering
    \begin{tabular}{l l l l l}
    \toprule
    Name & Symbol & Expression & Unit & Reference  \\
    \midrule
    Bulk thermal conductivity & $k_{\mathrm{b}}$ & $567/T$ & $\SI{}{W.m^{-1}.K^{-1}}$  & \citenum{Klinger1981}  \\
    Specific heat capacity & $c_{\mathrm{p}}$ & Equation~\ref{eq:cp_klinger} & $\SI{}{J.kg^{-1}}.K^{-1}$ & \citenum{Klinger1981}  \\
    Bulk density & $\rho_{\mathrm{b}}$ & 934 & $\SI{}{kg.m^{-3}}$  & SeaFreeze \\
    & &  &  &   \\
    Total thermal inertia & $\Gamma$ & Equation~\ref{eq:total_inertia} & $\SI{}{J.m^{-2}.K^{-1}.s^{-1/2}}$ & This study\\
    Low Thermal inertia Icy Moons & $\Gamma_{\mathrm{LTIM}}$ & 15 &  $\SI{}{J.m^{-2}.K^{-1}.s^{-1/2}}$ & Table \ref{tab:lowTI}  \\
    Contact thermal conductivity & $k_{\mathrm{con}}$ & Equation~\ref{eq:cond_sol_por}  &  $\SI{}{W.m^{-1}.K^{-1}}$ & Section~\ref{sec:porous_solid} \\
    Radiative thermal conductivity & $k_{\mathrm{rad}}$ & Equation~\ref{eq:cond_rad} & $\SI{}{W.m^{-1}.K^{-1}}$  & Section~\ref{sec:rad} \\
    Gas thermal conductivity & $k_{\mathrm{gas}}$ &  &  $\SI{}{W.m^{-1}.K^{-1}}$ & Section~\ref{sec:gas} \\

    Grain radius & $r_\mathrm{g}$ & $\SI{e-6}{} - \SI{e-3}{} $ & m &  \\
    Contact / Bond radius & $r_\mathrm{b}$ & Equation~\ref{eq:rb_express} & m &  \\
    Porosity & $\phi$ &$50 - 70\%$ &  & This study \\
    Density of porous $I_\mathrm{h}$ & $\rho$ &$ (1 - \phi) \rho_b$ &  $\SI{}{kg.m^{-3}}$ & By definition \\
    Surface tension & $\gamma$ & $0.17  e^{- {Q}/{RT}} $ & $\SI{}{W.m^{-2}}$ & \citenum{Jabaud2023}  \\
    Coordination number & $N_{\mathrm{C}}$ & Table \ref{tab:sphis} &  &  \\
    Emissivity & $\epsilon$ & 0.9 &  & \citenum{Spencer1999} \\
    Thermal skin depth & $\delta_{\mathrm{th}}$ & $\sqrt{{k P}/{\rho c\pi}}$ & m & Heat equation \\
    
    \bottomrule
    \end{tabular}
    \caption{Summary of the main parameters used and notations.}
    \label{tab:sumtab}
\end{table}

\section{Methods}
\label{sec:methods}
Thermal properties depend on thermodynamic conditions, especially temperature, material composition, and microstructure: grains geometry and their contacts, size distribution, sphericity, etc...
In this study, we explore the parameter space where the modeled thermal inertia $\Gamma$ does not exceed the estimated Low Thermal-inertia of Icy Moons $\Gamma_{\mathrm{LTIM}} = \SI{15}{}$, which we will refer hereafter as the ``$\mathit{\Gamma < \Gamma_{\mathrm{LTIM}}}$ \textit{relationship}''.

In a granular medium, heat transfer may occur via three modes: 1. Conduction through the grain contact areas in the solid phase, 2. Radiative conduction via photons emitted in the pore space and absorbed between neighboring grains, 3. Conduction through diffusion or convection of the gas phase in the pore space. 
The total conductivity $k$ combines these contributions:
\begin{equation}
    k = k_{\mathrm{\mathrm{con}}} + k_{\mathrm{\mathrm{rad}}} + k_{\mathrm{gas}}
    \label{eq:tot_cond}
\end{equation}
where $k_\mathrm{\mathrm{con}}$ is the contact conductivity of the porous medium, $k_{\mathrm{rad}}$ is its radiative conductivity and $k_{\mathrm{gas}}$ is the conductivity of the gas phase in the pores.
It follows from Equation \ref{eq:inertia}, that the total thermal inertia of the medium is expressed as
\begin{equation}
   \Gamma= \sqrt{\Gamma_\mathrm{\mathrm{con}}^2+\Gamma_{\mathrm{rad}}^2 + \Gamma_{\mathrm{gas}}^2}
    \label{eq:total_inertia}
\end{equation}

%you probably cannot use the same $\rho, c_p$ for the gas thermal inertia, but what about for radiative conductions?

Here, as in most icy moons thermal studies, we will assume that impurities, if present, have a negligible impact on the regolith thermal properties.
Hence, the thermal behavior of the regolith is primarily driven by that of crystalline water ice (Assumption \ref{as:cry_dom}).
A qualitative assessment of the effect of impurities and the implications of this assumption are discussed in Section \ref{sec:limits}.

\subsection{Contact conductivity in the solid phase}

\subsubsection{Bulk hexagonal water-ice}
\label{sec:bulk}

Thermal resistance theory \citep{Batchelor1977} tells us that the contact conductivity  $k_\mathrm{\mathrm{con}}$ is proportional to the bulk conductivity $k_b$ of the material it is made of 
\begin{equation}
    k_\mathrm{\mathrm{con}} \propto k_b.
\end{equation}
Thus, to model the contact conductivity, one must first estimate the bulk properties of hexagonal water ice under conditions typical of icy moons, effectively treating the thermal response of a smooth, solid ice slab.
For this purpose, we consider a temperature range of \SIrange{50}{175}{K}, representative of the observed lower and upper temperature limits on icy moons \citep{Spencer1987, Spencer1999, Howett2010}.

\begin{figure}[htpb]
    \centering
    \includegraphics[width=\linewidth]{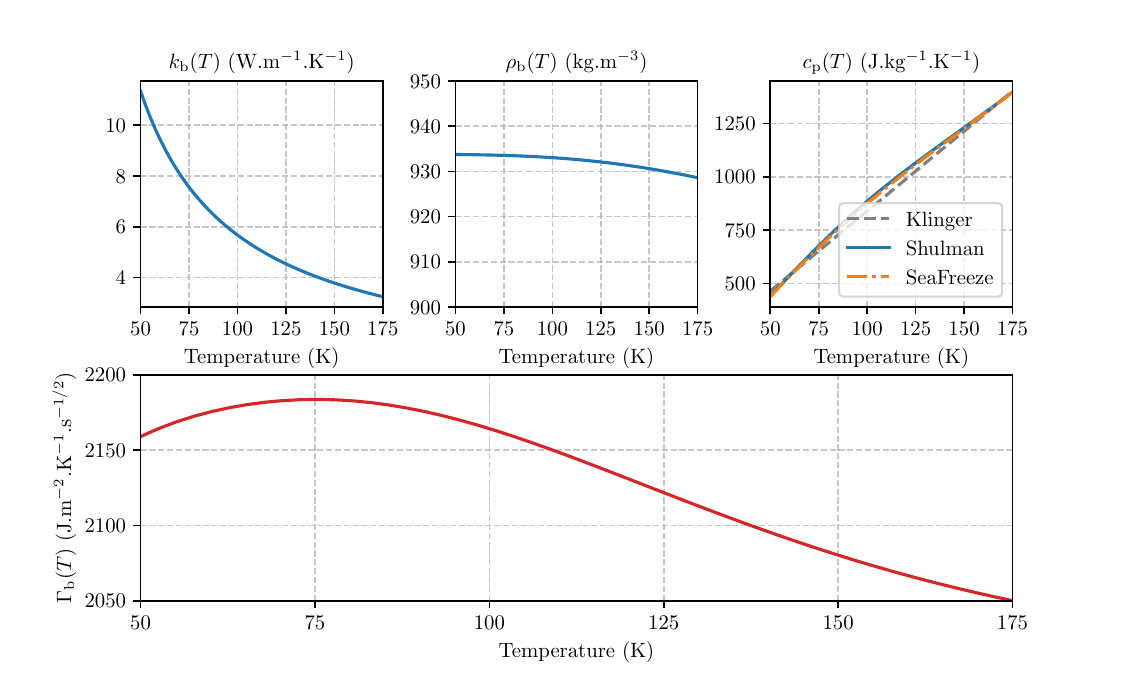}
    \caption{Bulk thermal properties of crystalline water ice for the typical range of temperature found on icy moons. \textit{Top, from left to right,} the bulk thermal conductivity, bulk density and specific heat capacity of hexagonal water ice $I_{\mathrm{h}}$. \textit{Bottom}, bulk thermal inertia of hexagonal water ice.}
    \label{fig:bulk}
\end{figure}

\paragraph{bulk conductivity}
Thermal conductivity of bulk hexagonal water ice is generally modeled using the law given by \cite{Klinger1981} $ k_{\mathrm{b}}(T) = {567}/{T}$, which as shown in Figure \ref{fig:bulk} decreases by an order of magnitude from 50K to 175K. 

\paragraph{bulk density}
Accurate values of hexagonal water ice density can be obtained by calling the SeaFreeze package \citep{Journaux2020} in the considered temperature range and low pressures of icy moons $\ll \SI{1}{Pa}$. 
Although the density of water ice is sometimes assumed to be around $\SI{918}{kg.m^{-3}}$, experimental studies \citep{Loerting2011, Satorre2018} pointed out that near the temperatures of icy moons, it is closer to $\SI{934}{kg.m^{-3}}$. In deed, density shows a slow decline as temperature increases, with  the $\SI{918}{kg.m^{-3}}$ value corresponding to the density at $\SI{0}{K}$ \citep{Loerting2011,Satorre2018}.
This trend is consistent with SeaFreeze computations (see Figure~\ref{fig:bulk}, \textit{middle}), which are based on the evaluation of Gibbs Local Basis Functions constructed to reproduce thermodynamic measurements of water ice phases. 
Thus, we note that for icy moons temperatures, assuming a constant $\rho_{\mathrm{b}} = \SI{934}{kg.m^{-3}}$ is a good approximation, as the density varies by less than 0.5\% in the $50-\SI{175}{K}$ temperature range.

\paragraph{Specific heat capacity}
Specific heat capacity increases significantly with temperature.  
Three derivations are compared in Figure~\ref{fig:bulk}: 
\begin{enumerate}

\item The simplest and most widely cited formulation is provided by \citet{Klinger1981}:
\begin{equation}
c_{\mathrm{p}}(T) = \SI{7.49}{} T + \SI{90}{J.kg^{-1}.K^{-1}}.
\label{eq:cp_klinger}
\end{equation}

\item A more accurate formulation, especially consistent with experimental data at lower temperature $T<\SI{100}{K}$, is given by \textcite{Shulman2004}:
\begin{equation}
\begin{split}
    c_{\mathrm{p}}(T) =& \SI{7.73}{} T ( 1 - e^{-\SI{1.263e-3}{} T^2}) \\ 
    &\times \left(1 + e^{-3\sqrt{T}} \times \SI{8.47e-3}{} T^6 + \SI{2.0825e-7}{} T^4 e^{-\SI{4.97e-2}{T}} \right)
\end{split}
\end{equation}

\item Thermodynamically computed values from the SeaFreeze package are also included \citep{Journaux2020}.
\end{enumerate}

Within the temperature range relevant to icy moons, these three formulations produce nearly identical values, with the \textcite{Shulman2004} and SeaFreeze values showing the strongest agreement. 
The approximation by \textcite{Klinger1981} deviates by at most $\pm 6\%$ from the more precise values, reducing to $\pm 3\%$ when propagated to thermal inertia calculations. 
While this approximation is suitable for analytical purposes, to be as accurate as possible, we adopt the \textcite{Shulman2004} formula for the remainder of this study.

The resulting bulk thermal inertia of crystalline water ice can be derived using Equation~\eqref{eq:inertia} and temperature-dependent expressions for $k_{\mathrm{b}}(T)$, $\rho_{\mathrm{b}}(T)$, and $c_{\mathrm{p}}(T)$. As shown in Figure~\ref{fig:bulk} (\textit{Bottom}), the bulk thermal inertia varies from \SI{2813}{} to \SI{2050}{} across the surface temperature range of icy moons, showing a general decrease of about 6\% over this range. These variations will directly impact the effective thermal inertia of the porous regolith (as shown later in Section \ref{sec:porous_solid}).

To estimate the influence of these bulk properties on remote sensing measurement, let's consider a typical region on Europa, where nighttime temperatures are near \SI{75}{K} and noon temperatures near \SI{125}{K} \citep{Mergny2024heatcode}. 
Based on Figure~\ref{fig:bulk}, the day-night variation of thermal inertia in these conditions would be around 3\%. 
While more complex models could improve accuracy by accounting for temperature-dependent bulk properties, the current uncertainties in Table~\ref{tab:lowTI} ($\Delta\Gamma_{\mathrm{LTIM}}/\Gamma_{\mathrm{LTIM}}\sim \pm 30\%$) are significantly larger than 6\%. 
However, bulk properties are not the only temperature-sensitive parameters. As it will be demonstrated in Section~\ref{sec:rad_temperature}, radiative conduction shows a strong temperature dependence, leading to more drastic variations in thermal inertia with temperature.

\subsection{Porous water-ice}
\label{sec:porous_solid}

We consider the regolith to be a porous medium made of spherical grains of uniform size surrounded by pores at low pressure (see discussion of these assumptions in Section \ref{sec:limits}). 
The porosity $\phi$ is defined as the volume occupied by the pore spaces divided by the total volume. 
This heat transfer in the solid phase passes from grain to grain through their areas of contact.

\begin{figure}[htbp]
    \centering
    \includegraphics{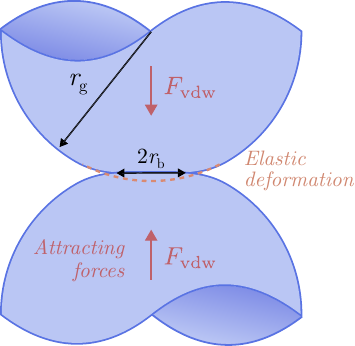}
    \caption{Representation of how two grains of radius $r_\mathrm{g}$ form an initial contact radius $r_\mathrm{b}$ based on Hertz theory. The driving forces are due to van der Waals interactions here denoted as $F_{\mathrm{vdw}}$.}
    \label{fig:hertz-contact}
\end{figure}

\begin{figure}[htbp]
    \centering
    \includegraphics{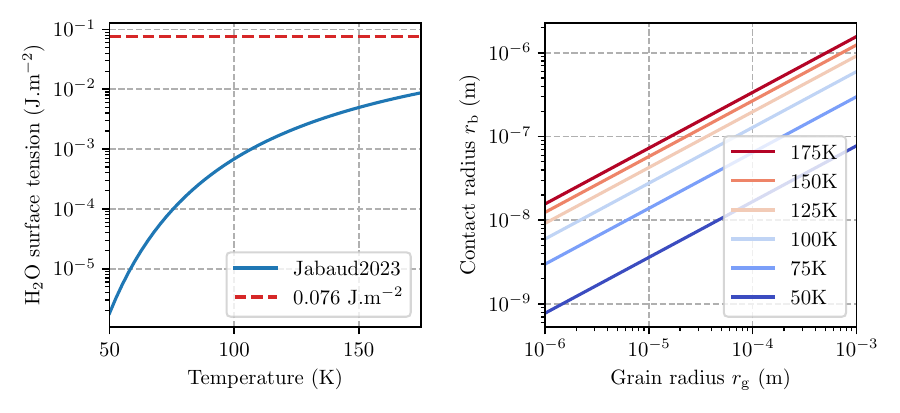}
    \caption{(\textit{Left}) Surface tension from \citet{Jabaud2023} as function of temperature. (\textit{Right}) Initial contact radius as function of grain radius and temperature. Note that ice sintering may increase the contact radius as discussed in Section~\ref{sec:sintering}.}
    \label{fig:contact-evo}
\end{figure}

When two grains are close to each other, they naturally exert an attractive force bringing them into contact. 
This attractive force then deforms the grains elastically (see Figure \ref{fig:hertz-contact}, dotted sphere contours). 
This results in a very small contact region that allows heat to pass through the solid phase.
The Hertzian theory \citep{Hertz1882} gives the radius $r_{\mathrm{b}}$ of the contact area based on the external force experienced by the grains, here of same size and properties:
\begin{equation}
    r_{\mathrm{b}}^3 = \dfrac{3}{4} \dfrac{1-\nu^2}{E(T)} r_{\mathrm{g}} F(r_{\mathrm{g}}, T)
    \label{eq:hertz_bond}
\end{equation}
where $E(T) = \SI{6.6e9}{} (4.276 - 0.012 \,T) \SI{}{Pa}$ \citep{Ferrari2016} is the ice elastic modulus, $\nu=0.33$ its Poisson ratio \citep{Ferrari2016}, and $F(r_{\mathrm{g}}, T)$ is the force that acts on the spheres.
For sub-millimeter grains of water ice under low gravity, the van der Waals force is orders of magnitude higher than their weight, when no external load is applied \citep{Gundlach2012, Ferrari2016}. Such van der Waals force can be obtained from the theory of \citet{Johnson1971} with the expression
\begin{equation}
    F(r_{\mathrm{g}}, T) = 3 \pi \gamma(T) r_{\mathrm{g}}
\end{equation}
where $\gamma(T)$ is the surface tension of water ice. 
The surface tension of water ice has been widely debated in the literature, with earlier studies \citep{Gundlach2012, Ferrari2016, Molaro2019} typically adopting a constant value near $\gamma \sim 0.06-\SI{0.076}{J.m^{-2}}$. 
However, recent experimental work within icy moons temperature range suggests that the surface tension is likely much lower. 
Specifically, \textcite{Jabaud2023} demonstrated that surface tension decreases drastically at lower temperatures, following an Arrhenius-type dependence.
\begin{equation}
    \gamma(T) = 0.17 \exp{\left( -\dfrac{Q}{R_{\mathrm{gas}}T} \right)}
\end{equation}
with the activate energy $Q = 4.6 \pm \SI{0.3}{kJ.mol^{-1}}$ and $R_{\mathrm{gas}}$ the gas constant. 
The variation of surface tension across the temperature range considered in this study is shown in Figure~\ref{fig:contact-evo} (\textit{Left}). These new experimental values reveal a drastic decline in surface tension by nearly four orders of magnitude. Even at the highest temperature considered here (\SI{175}{K}), the surface tension remains an order of magnitude lower than the typical values previously reported in the literature.
The accuracy of this parameter strongly influences contact conductivity models, as it directly governs the contact region radius and, consequently, the efficiency of heat transfer.

Substituting the van der Waals force expression into Equation \eqref{eq:hertz_bond}, leads to the radius of the initial contact area, \textit{i.e.} the bond radius:
\begin{equation}
    r_{\mathrm{b}}(T, r_{\mathrm{g}}) =   \sqrt[3]{\dfrac{9}{4} \, \dfrac{1-\nu^2}{E(T)} \, \pi \gamma(T) r_{\mathrm{g}}^2 }.
    \label{eq:rb_express}
\end{equation}
We show in Appendix~\ref{app:molarosint} how this expression is equivalent to what \citet{Ashby1974} and \citet{Molaro2019} describe as ``\textit{Stage 0}'' sintering.
The contact radius $r_{\mathrm{b}}$ is shown as a function of grain radius and temperature in Figure~\ref{fig:contact-evo} (\textit{Right}).
Due to the exceptionally low surface tension values at low temperatures, near $\SI{50}{K}$, the bond radius can decrease to as little as $\SI{1}{nm}$ for micrometer-size grains. Such small bond radii may not be relevant so their potential impact on our results at cold temperatures is discussed in Section~\ref{sec:limits}.
Ice sintering can increase the contact radius, and would be significant at higher temperatures (see Section~\ref{sec:sintering}).

\textcite{Chan1973} conducted a thermal conductivity study of packed spheres based on the Hertz factor, defined as $H \equiv r_{\mathrm{b}} / r_{\mathrm{g}}$ \citep{Gundlach2012}, where $r_{\mathrm{b}}$ is the contact area radius (referred to as the \textit{bond radius} hereafter). Following \textcite{Ferrari2016}, the general formula for contact conductivity is expressed as:
\begin{equation}
    k_{\mathrm{\mathrm{con}}} = k_{\mathrm{b}}(T) \, S(\phi) \, \dfrac{r_\mathrm{b}}{r_{\mathrm{g}}},
    \label{eq:cond_sol_por}
\end{equation}
where $S(\phi)$ is a porosity function accounting for porosity effects, and the Hertz factor is directly incorporated.

Table~\ref{tab:sphis} lists the most common contact conductivity models from the literature. Although these models are formulated differently in their original publications, they can all be reduced to the general form of Equation~\eqref{eq:cond_sol_por} (see Appendix~\ref{app:cond_express}). 
All models \citep{Adams1993, Shoshany2002, Lehning2002b, Gusarov2003, Gundlach2012, Mergny2024h} satisfy the physical limit $\lim_{\phi \to 1} S(\phi) = 0$, ensuring conductivity approaches zero at 100\% porosity. 
Also, for all models, Equation~\eqref{eq:cond_sol_por} fails to reproduce the expected limit $\lim_{\phi \to 0} k = k_b$ for very low porosity although this limit is never reached for an icy moon's regolith. 
We note that at high porosity, Adams\&Sato1993 and SNOWPACK models have a coordination number that becomes unrealistic, as shown in Appendix \ref{app:coord}.

\begin{table}[htbp]
\centering

    \begin{tabular}{l l l l}
        \toprule
        {Model} & ${S(\phi)}$ & {Additional parameters}  & Also used by\\
        \midrule
        Gundlach2012 & $f_1 \exp{(f_2 (1-\phi))} $&  $f_1 = \SI{5.18e-2}{}$, $ f_2= \SI{5.26}{}$ &  \citenum{Chan1973} \\
        \addlinespace[0.3cm] % Adds 0.5cm space after this row
        
        SNOWPACK & $ \sqrt[3]{\dfrac{3}{4\pi}(1-\phi)} \dfrac{\pi^2}{32} N_{\mathrm{c}}(\phi) $ &  $N_{\mathrm{C}}^{*}(\phi) = \sum\limits_{i=0}^{5} a_i^{\dag} \rho_b (1-\phi)^i$  & \\
        \addlinespace[0.3cm] % Adds 0.5cm space after this row
        
        Shoshany2002 & $(1-\phi/\phi_c)^{\alpha(\phi)^{n^{**}}}$ & $\alpha(\phi) = 4.1 \phi + 0.22$, $ \phi_c = 0.7$ & \citenum{Raza2026}\\
        \addlinespace[0.3cm] % Adds 0.5cm space after this row
        
        Gusarov2003
        & $\dfrac{1}{\pi}(1-\phi) N_{\mathrm{C}}(\phi) $&  $N_{\mathrm{C}}^{*} (\phi) = 2.17 e^{\rho_b (1-\phi)\times \SI{1.9e-3}{}}$ &  \citenum{Kleer2021, Camarca2023, Thelen2024, Lange2026} \\
        \addlinespace[0.3cm] % Adds 0.5cm space after this row
        
        Adams\&Sato1993 & $\dfrac{1}{\pi}(1-\phi) N_{\mathrm{C}}(\phi) $ &  $N_{\mathrm{C}}^{*} (\phi) = \sum\limits_{i=0}^{2} b_i^{\ddag} \rho_b (1-\phi)^i$ & \\
        \addlinespace[0.3cm] % Adds 0.5cm space after this row
        
        MultIHeaTS & $(1-\phi)$ & & \citenum{Mergny2024i} \\
        \addlinespace[0.3cm] % Adds 0.5cm space after this row
        \bottomrule
    \end{tabular}

    \begin{threeparttable}
    \begin{tablenotes}
    \footnotesize
        \item[*]  $N_{\mathrm{C}}$ is the coordination number, \textit{i.e.} the number of contacts each grain has.
        \item[**] $n$ is the pore size distribution in \citet{Shoshany2002}. Here with $n=3$, as in \citet{Raza2026}.
        \item[$\dag$] $a_i$ given in \citet{Lehning2002b}, their equation 17.
        \item[$\ddag$] $b_i$ given in \citet{Adams1993}, their equation 1.   
    \end{tablenotes}
    \end{threeparttable}
    
    \caption{List of thermal conductivity contact models for porous media in planetary science literature, as per Equation~\ref{eq:cond_sol_por}. Thermal inversion studies using these models are also indicated.
    }
  \label{tab:sphis}
\end{table}
%I should cite Wood2020 somewhere and probably use his incredible (but complex) model to compare with others!

\begin{figure}[htbp]
    \centering
    \hspace{2cm}
    \includegraphics{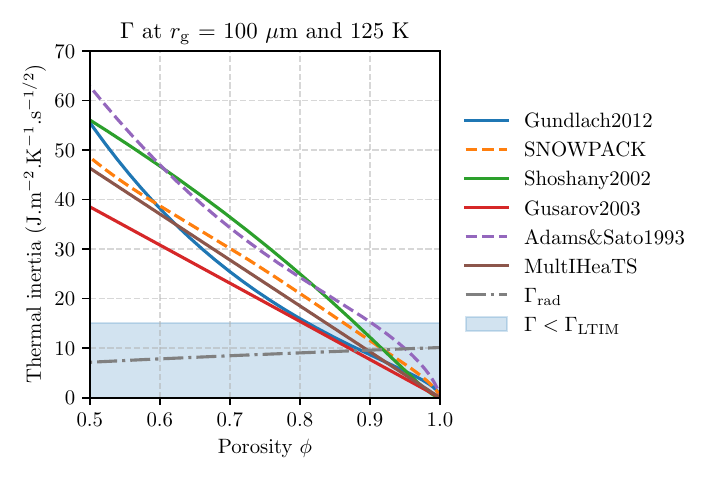}
    \caption{Thermal inertia of most common porous ice conductivity models at $\SI{125}{K}$ for grain radius $r_{\mathrm{g}} = \SI{100}{\micro m}$ as function of porosity. The thermal inertia due to the radiative contribution is also represented. For a detailed exploration of the grain radii and temperature dependency, please refer to Figure \ref{fig:param_allowed}. Note that the Adams\&Sato1993 and SNOWPACK models may be invalid above 80\% porosities due to unrealistic coordination number values (Appendix~\ref{app:coord}) and that Shoshany2002 model is plotted for $n=3$ following \citet{Raza2026}.
    }
    \label{fig:all_TIs}
\end{figure}

The thermal inertia due to conduction by contact between grains, $\Gamma_\mathrm{\mathrm{con}}$, is then derived from Equation~\eqref{eq:inertia} by choosing a model for $k_\mathrm{\mathrm{con}}$, the porous medium density $\rho$, and the specific heat capacity of water ice $c_p$. By definition, the density of the porous medium is given by:
\begin{equation}
    \rho(\phi) = (1 - \phi) \rho_b,
    \label{eq:density}
\end{equation}
where the gas phase density is neglected, as it is orders of magnitude lower than that of the solid phase. For a given grain radius and temperature, the bond radius is calculated using Equation~\eqref{eq:rb_express} and then integrated to compute the contact conductivity $k_\mathrm{\mathrm{con}}$ for any porosity, using any of the models listed in Table~\ref{tab:sphis}.

Figure \ref{fig:all_TIs} presents the computed thermal inertia for all the contact conductivity models within their domain of validity (that is, excluding low porosity). 
Despite coming from different approaches and fields (e.g., planetary science, cryosphere science), these models show good agreement, particularly at high porosity, which as discussed later, is most representative of icy moon regolith.
This implies that the choice of model from Table~\ref{tab:sphis} and Equation~\eqref{eq:cond_sol_por} will not have a significant impact on the computed thermal inertia, making our results more robust to such arbitrary choice.
While the MultIHeaTS conductivity expression is rather simple and not experimentally validated \citep{Mergny2024h}, it aligns well with more established models. For applications that prioritize simplicity over high precision, this model offers a practical alternative while maintaining good agreement with robust formulations.

The thermal inertia from contact conductivity can be rewritten as
\begin{equation}
    \Gamma_\mathrm{\mathrm{con}}   = \Gamma_{b} \sqrt{(1-\phi)S(\phi) H(r_{\mathrm{b}}, r_{\mathrm{g} })}.
    \label{eq:Isolid}
\end{equation}
This formulation highlights that variations in porous thermal inertia directly scale with the bulk thermal inertia (Figure~\ref{fig:bulk}).
The reader may refer to Section \ref{sec:bulk} to estimate the impact of temperature variations on the contact thermal inertia of the porous regolith.

\subsection{Radiative conductivity}
\label{sec:rad}
The second mechanism of heat conduction in porous media is radiative conduction within the pore space. Grains emit photons via gray-body radiation, which travel through the pore space (unaffected by the gas due to low pressure) until absorption by neighboring grains. 
%This process is possible because water ice is opaque at the wavelengths it emits (is it not always the case?)
While radiative conduction is often neglected due to the low temperatures of icy moons, since contact conductivity is greatly reduced at high porosity and instead radiative transport increases with porosity, it may contribute significantly to heat transport.

\textcite{Dullien1991} derived an expression for radiative conductivity, $k_{\mathrm{rad}}$, based on the mean free path of photons in a random porous medium:
\begin{equation}
    k_{\mathrm{\mathrm{rad}}} = 8 \sigma_{\mathrm{SB}} \epsilon e_1 \dfrac{\phi}{1 - \phi} r_{\mathrm{g}} T^3,
    \label{eq:cond_rad}
\end{equation}
where $\sigma_{\mathrm{SB}}$ is the is the Stefan–Boltzmann constant, $\epsilon$ is the emissivity of the material and $e_1 = 1.34$ is a fitted parameter from \textcite{Gundlach2012}. 

We note that alternative radiative conduction expressions exist, such as the one from \citet{Breitbach1980}. \citet{Ferrari2016} compared the two in their figure 1 and concluded that they lead t relatively similar predictions, although \citet{Gundlach2012} radiation model is more sensitive to porosity. We explore how choosing Breilach's model would influence our results in Appendix~\ref{sec:BB_model}.

The radiative conductivity depends on porosity, grain radius, and, most critically, scales with the cube of the medium's temperature.
Using Equation~\eqref{eq:inertia}, the radiative thermal inertia can be expressed as:
\begin{equation}
    \Gamma_{\mathrm{rad}} = \sqrt{8 \sigma_{\mathrm{SB}} \epsilon e_1 \rho_b \phi r_{\mathrm{g}} c_p(T) T^3},
    \label{eq:Irad}
\end{equation}
and is plotted alongside the contact conductivity models in Figure~\ref{fig:all_TIs}.
It shows how for grain radii $\geq \SI{100}{\micro m}$ radiative thermal inertia can become significant, if not dominant, compared to contact thermal inertia.

\subsection{Gas conductivity}
\label{sec:gas}

To estimate the importance of gas conductivity, we follow \citet{Piqueux2009} and compute the Knudsen number, defined as the ratio of the mean free path of \ce{H2O} molecules to the typical pore size $l_\mathrm{pore}$:
\begin{equation}
    \mathrm{Kn} = \dfrac{k_\mathrm{b} T}{ \sqrt{2} \pi P_{\mathrm{gas}} \, d_{\ce{H2O}} \, l_\mathrm{pore}},
\end{equation}
where $k_\mathrm{b}$ is Boltzmann's constant, $d_{\ce{H2O}} = \SI{2.75e-10}{m}$ is the diameter of a water molecule, and $P_{\mathrm{gas}}$ is the gas pressure. The typical pore size is obtained from \citet{Piqueux2009}:
\begin{equation}
    l_\mathrm{pore} = \sqrt[3]{\dfrac{\phi}{1-\phi}} \times r_\mathrm{g}.
\end{equation}
Using typical values for icy moons, $T = \SI{100}{K}$, $r_\mathrm{g} = \SI{100e-6}{m}$, and $P_{\mathrm{gas}} \approx \SI{e6}{Pa}$ \citep[exosphere pressure on Europa]{Hall1995}, we find $\mathrm{Kn} > \SI{e7}{}$. Since large Knudsen numbers ($\mathrm{Kn} > 10$) correspond to the free molecular regime, gas conductivity is greatly reduced and negligible compared to other heat transport modes \citep{Piqueux2009}. Given that $\mathrm{Kn}$ is multiple orders of magnitude above 10 in our typical example, we estimate that for all realistic icy moon conditions gas conduction cannot be an effective mode of heat transport in the regolith (\textit{i.e. }$k_{\mathrm{gas}} \ll k_{\mathrm{con}}, k_{\mathrm{rad}})$.

%\citet{Wasilewski2021} says " gas transport is negligible, however it is important to note that within systems where pressures and particle diameters lead to low Knudsen numbers, gas conductivity may be dominant"
%Gundlach2012, also neglect gas transport because of the low pressure in the experiment.
%But that would be interesting to look into it \citep{Davidsson2002}!
%relevant gas densities would also lower the mean free path of the photons assumed in the Radiative conductivity Section.

\section{Results}

\subsection{Time varying thermal inertia}
\label{sec:rad_temperature}
%Piqueux2011: ] We have shown that the thermal inertia of the regolith may double over the range of the Martian temperatures.
%and Such variations indicate that the specific heat is the most temperature‐
%Piqueux2009a: for composition: has sug- gested that the nature of the solid composing the particles of a natural surface does not influence the bulk thermal conduc- tivity of a soil because regardless of its composition, the heat will propagate significantly faster there than in the intergrains regions and gas.

On a study of Martian soil, \citet{Piqueux2011} showed that the thermal inertia of the regolith may double over the range of Martian temperatures.
While under icy moon conditions, gas conduction is not as effective as on Mars, radiative heat transport has a strong temperature dependency that may also lead to time variations during diurnal or seasonal cycles.

\begin{figure}[htbp]
    \centering
    \includegraphics{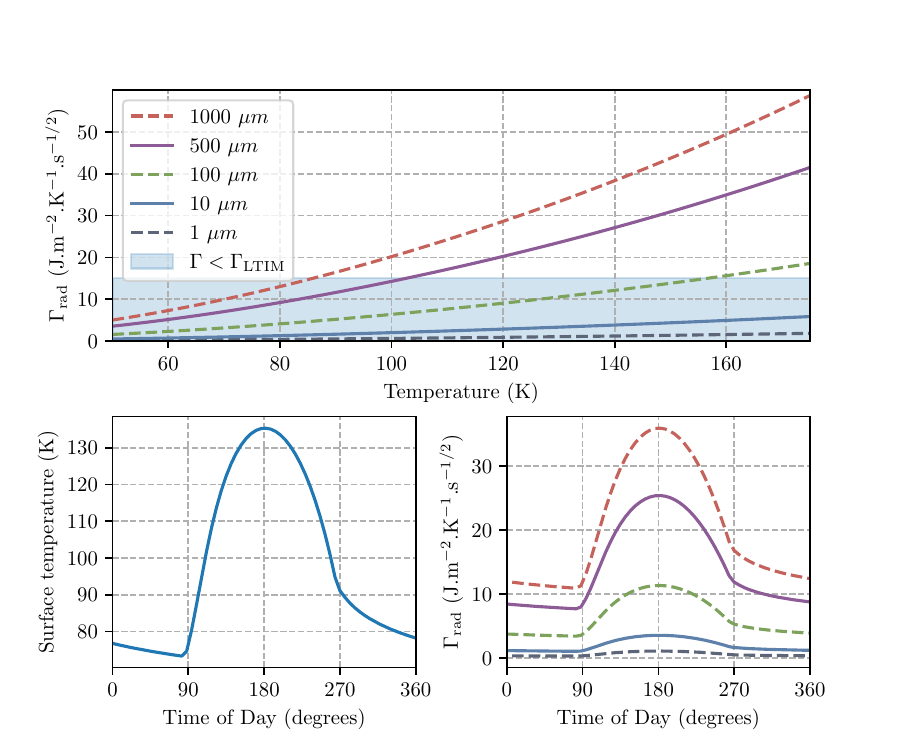}
    \caption{(Top) Radiative thermal inertia as function of temperature for different grain radii and a porosity of $\phi = 75\%$. (Bottom Left) Simulated Europa's surface temperature at the equator for an albedo of 0.6. (Bottom Right) How would radiative thermal inertia change throughout Europa's day cycle, the same legend as the top plot applies.}
    \label{fig:rad_temp}
\end{figure}

Figure~\ref{fig:rad_temp} (\textit{Top}) shows the radiative thermal inertia for a regolith with 75\% porosity across the temperature and grain radius ranges typical of icy moons.
While this choice of porosity may seem arbitrary, Equation~\eqref{eq:Irad} reveals that radiative thermal inertia depends only weakly on porosity, scaling as $\Gamma_{\mathrm{rad}} \propto \sqrt{\phi}$. 
Given realistic porosities in the range $\phi \in [0.5, 1]$, the ratio of radiative thermal inertias for two porosities $\phi_1$ and $\phi_2$ is:
\begin{equation}
    \dfrac{\Gamma_{\mathrm{rad}}(\phi_1)}{\Gamma_{\mathrm{rad}}(\phi_2)} = \sqrt{\dfrac{\phi_1}{\phi_2}}.
    \label{eq:rad_var_phi}
\end{equation}
which for the two end member values of porosity ($\phi_1=0.5$ and $\phi_2\to1$) leads to a radiative thermal inertia that varies by only $\pm 10\%$.

While Equation~\eqref{eq:inertia} also shows a square root dependency on grain radii, the wide range of possible grain radii, spanning three orders of magnitude ($r_{\mathrm{g}} \in [\SI{1}{\micro m}, \SI{1000}{\micro m}]$), means their influence on radiative thermal inertia is far more significant. 
As delimited by the blue region in Figure~\ref{fig:rad_temp}, grains larger than $\SI{500}{\micro m}$ to \SI{1}{mm} rapidly make radiative heat transfer too efficient compared to what we expect from thermal measurements made on icy moons.

The strongest dependency, however, is on temperature. 
A first look at Figure (\ref{fig:rad_temp}, \textit{Top}) shows that grains of radius $\SI{500}{\mu m}$ and larger are not compatible with the $\Gamma < \Gamma_{\mathrm{LTIM}}=15$ relationship as soon as temperature exceed $\SI{100}{K}$. 
Since the specific heat capacity varies linearly with temperature ($c_p(T) \propto T$), the radiative thermal inertia scales as $\Gamma_{\mathrm{rad}} \propto T^2$. Similar dependencies have been previously derived by \textcite{Lellouch2013}, with only minor differences due to alternative formulations for the specific heat capacity.
Over the temperature range $T \in [\SI{50}{K}, \SI{175}{K}]$, this leads to an increase in thermal inertia of slightly more than one order of magnitude, as described by the relation:
\begin{equation}
    \dfrac{\Gamma_{\mathrm{rad}}(T_1)}{\Gamma_{\mathrm{rad}}(T_2)} = \left(\dfrac{T_1}{T_2}\right)^2,
    \label{eq:rad_var_temp}
\end{equation}
where $T_1$ and $T_2$ are the two temperature endmembers being compared.
To illustrate this behavior, we consider a typical day-night cycle on Europa's equator as shown in Figure~\ref{fig:rad_temp} (\textit{Lower Left}).
In such a region, temperatures vary from \SI{75}{K} at the end of the night to \SI{135}{K} near noon. This means that the efficiency of radiative conduction varies significantly over the course of a day.
Using Equation~\eqref{eq:rad_var_temp}, we compute that $    \Gamma_{\mathrm{rad}}(\SI{135}{K}) = 3.24 \times \Gamma_{\mathrm{rad}}(\SI{75}{K})$, meaning that the radiative thermal inertia at noon is more than three times greater than at the end of the night, regardless of porosity or grain size.
The same surface measured at night may have very different thermal inertia than during the day \citep{Wood2020}.
This variation is illustrated in Figure~\ref{fig:rad_temp} (\textit{Right}), showing how radiative thermal inertia changes over such a temperature cycle.
However, since most observations probe multiple centimeters deep (see Table~\ref{tab:lowTI}), temperature variations would be less drastic.
Still, this effect is never taken into account when inverting thermal data acquired at different local hours, while it could change the estimated values of thermal inertia. As \citet{Ferrari2016} pointed out, considering the significant temperature dependency of thermal inertia, it should be systematically included in the thermal models.

\subsection{Maximum grain radius}

\begin{figure}[htbp]
    \centering
    \includegraphics{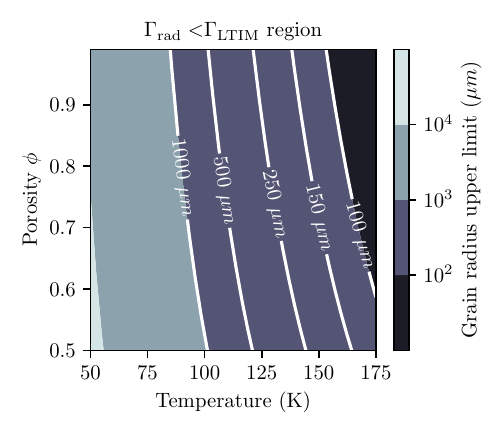}
    \caption{Maximum grain radius allowed for radiative heat transfer to respect the $\Gamma < \Gamma_{\mathrm{LTIM}}=15$ relationship, shown as a function of porosity and the temperature range of icy moons.}
    \label{fig:uppergrain}
\end{figure}
We first turn off contact conduction to determine if radiative conduction alone could become significant and constrain the surface properties.
To quantify how the $\Gamma_{\mathrm{rad}} < \Gamma_{\mathrm{LTIM}}$ relationship gives constraints on the regolith microstructure, we derive from Equation~\eqref{eq:cond_rad} an upper bound on grain radius for any given temperature and porosity:
\begin{equation}
   r_{\mathrm{g}} < \dfrac{\Gamma_{\mathrm{LTIM}}^2}{8\sigma_{\mathrm{SB}} \epsilon e_1 \rho_b \phi c_p(T) T^3}. 
   \label{eq:rgrad}
\end{equation}
This inequality defines the maximum allowable grain radius for a surface with thermal inertia $\Gamma_{\mathrm{LTIM}}$ at temperature $T$ and porosity $\phi$.

Figure~\ref{fig:uppergrain} illustrates how this constraint evolves over the typical porosity and temperature ranges of icy moons. 
Some key observations include:
1. As predicted by Equation~\eqref{eq:rad_var_phi}, radiative thermal inertia in \citet{Gundlach2012} formulation only slightly varies with porosity across the range $\phi \in [0.5, 1]$.
2. For surfaces with temperatures above \SI{100}{K}, typical of the Galilean icy moons, grain radii must remain below \SI{1}{mm} to ensure $\Gamma < \Gamma_{\mathrm{LTIM}}$.
3. Higher surface temperatures, such as those on Callisto, the warmest regions of Ganymede and Europa, or near Enceladus’ Tiger Stripes, impose stricter limits. Grain radii cannot exceed $\SI{200}{\micro m}$ to maintain thermal inertia below $\Gamma < \SI{15}{}$. For $T > \SI{160}{K}$, computations show that the maximum allowable grain radius for all reasonable porosities is $r_{\mathrm{g}} < \SI{150}{\micro m}$.

\subsection{Constraining porosity and grain radius}

\begin{figure}[htbp]
    \centering
    \includegraphics{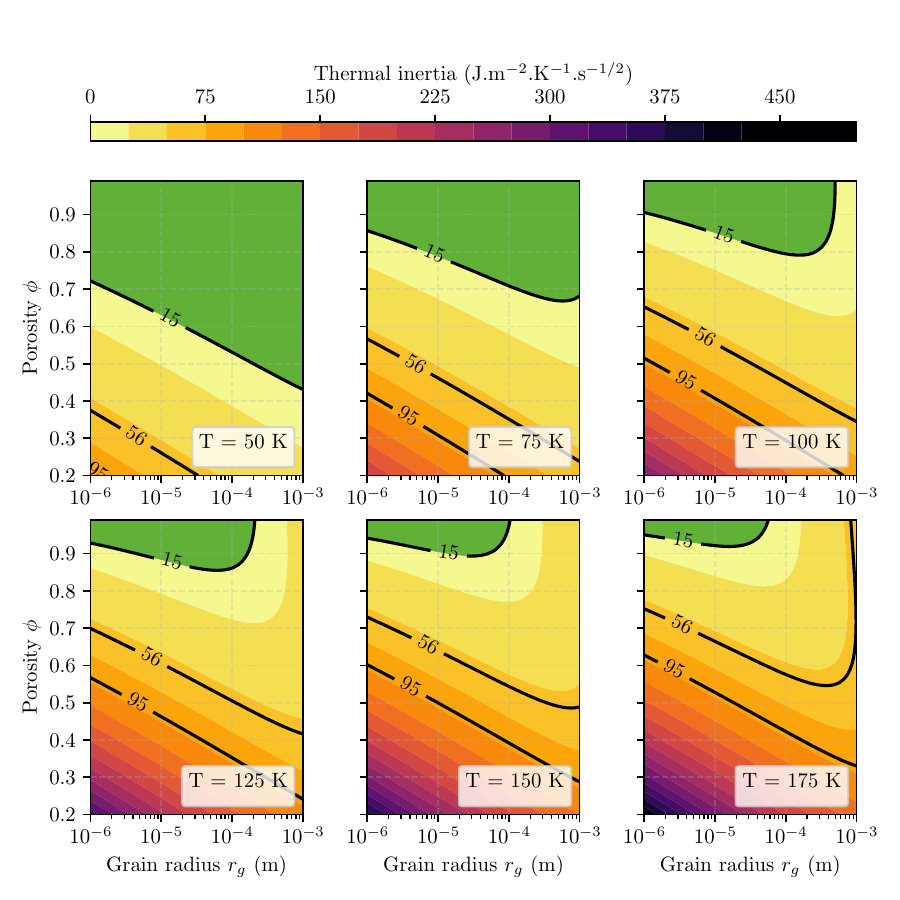}
    \caption{Thermal inertia of porous water ice as function of porosity, grain radius and temperature. The green area shows the range of allowed porosities and grain radiis to respect the $\Gamma < \Gamma_{\mathrm{LTIM}}$ relationship. This Figure applies to all icy moons, but for reference we have also shown the mean thermal inertia of Europa derived from Galileo PPR $\Gamma = 56$ \citep{Lange2026} and the mean value from ALMA (Band 6), $\Gamma = 95$ \citep{Trumbo2018}, represented by two black contours.
    }
    \label{fig:param_allowed}
\end{figure}

Since radiative conduction and contact conduction may contribute comparably to the total thermal inertia of the regolith, it is logic to now consider both contributions simultaneously. 
Any of the models listed in Table~\ref{tab:sphis} can be used to compute the contact conductivity. 
For this section, we use Gundlach2012 model, though the choice has minimal impact on results as discussed in Section \ref{sec:limits}, as all models produce similar values at high porosity (Figure~\ref{fig:all_TIs}).
Combining the expressions for $\Gamma_\mathrm{\mathrm{con}}$ \eqref{eq:Isolid} and $\Gamma_{\mathrm{rad}}$ \eqref{eq:Irad} with Equation~\eqref{eq:total_inertia} gives the expression of the regolith's thermal inertia:
\begin{equation}
    \Gamma  (\phi, r_{\mathrm{g}}, T) = \sqrt{ k_b(T) c_p(T) S(\phi)(1-\phi) r_{\mathrm{b}}(T)/r_{\mathrm{g}} + 8 \sigma_{\mathrm{SB}} \epsilon e_1 \rho_b \phi r_{\mathrm{g}} c_p(T) T^3}.
    \label{eq:Itot}
\end{equation}
This expression reveals that the thermal inertia depends primarily on three parameters: the porosity, grain radius, and temperature. 
In Figure \ref{fig:param_allowed} is plotted the total thermal inertia $\Gamma$ in this three parameters space for their relevant range of values on icy moons.

A key observation from Figure \ref{fig:param_allowed} is that a single thermal inertia value never corresponds to a unique pair of porosity and grain radius. 
Tracing the contour line of a given thermal inertia (\textit{e.g.}, the $\Gamma = \SI{15}{}$, \textit{black contour}) reveals all possible combinations of grain radii and porosities that give that same thermal inertia. 
This highlights the degeneracy inherent in using a single parameter like thermal inertia to characterize the thermal response of planetary surfaces, an important issue that will be addressed in detail in a forthcoming article.

Following our previous approach, we examine the parameter space that satisfies the relationship $\Gamma < \Gamma_{\mathrm{LTIM}} = \SI{15}{}$, represented by the green regions in Figure~\ref{fig:param_allowed}. Our analysis reveals that achieving such LTIM imposes strict constraints on porosity and grain radius.

Any surface with temperatures $\geq \SI{100}{K}$ during thermal measurements must show extremely high porosity ($>80\%$) and small grain radii $< \SI{500}{\micro m}$. 
Failure to meet these constraints would result in a higher effective thermal inertia than what was currently inferred by the literature from space observations. 
This temperature range is characteristic of the Galilean moons, implying that the regolith of Europa, Ganymede, and Callisto must, at minimum, adhere to these constraints if their estimated $\Gamma_{\mathrm{LTIM}}$ is accurate.

For Ganymede and Callisto, where surface temperatures can exceed \SI{150}{K}, even higher porosities ($>90\%$) and smaller grain radii ($< \SI{200}{\micro m}$) are required to maintain the $\Gamma<\Gamma_{\mathrm{LTIM}}$ relationship.
Enceladus’ Tiger Stripes, with the highest recorded temperatures on any icy moon, face the tightest constraints: if a thermal inertia as low as \SI{15}{} would be estimated, the required porosity would have to exceed $92\%$, with grain radii $< \SI{50}{\micro m}$.

Very low temperatures ($\SI{50}{K}$) is the only case that has considerably relaxed constraints because 1) radiative conduction is not efficient at these temperatures, and 2) the contact radius $r_{\mathrm{b}}$ is abnormally small (resulting from low surface tension, as discussed in Section~\ref{sec:porous_solid}). 
Even under these loosest constraints, porosities cannot drop below $40\%$.

Finally, we note that Figure~\ref{fig:param_allowed} is not specific to the case of $\Gamma < \SI{15}{}$, but can be used to compare with any thermal inertia $\Gamma_{\mathrm{LTIM}}$. Thus, other studies can use this figure and its color bar to estimate, under our modeling assumptions, the allowed porosity and grain radii ranges for an icy regolith to satisfy $\Gamma < \Gamma_{\mathrm{LTIM}}$.
For example, Figure~\ref{fig:param_allowed} shows the contour line $\Gamma = 56$, the median thermal inertia derived from recently revisited Galileo PPR observations of Europa \citep{Lange2026}, and the $\Gamma = \SI{95}{}$ contour line, the average thermal inertia derived from ALMA band 6 observations of Europa \citep{Trumbo2018}. These observations of the same icy moon, probing different depths, lead to different porosity/grain radius ranges, suggesting a layering in the regolith as will be discussed in Section~\ref{sec:layering}.
%Thelen2024: 56–184

\subsection{Comparison with the literature}
\label{sec:lit}

Overall our thermal analysis showed that the range of porosities compatible with the thermal measurement made on all icy moons is narrowed in the extremely high porosity values, from $>80\%$ to $90\%$. Are these results coherent with other approaches in the literature?

\paragraph{Photometry}

Photometry is the study the light scattered by a surface relative to the geometry of observation.
Photometric studies provide insights into the physical properties and microstructure of the surface, such as compaction, shape, roughness, transparency, etc...  \citep{Belgacemthesis}
Since the late 1980s, photometric analysis combining Earth-based observations and Voyager images have suggested that Europa's regolith porosity to be close to $96\%$ porosity \citep{Buratti1983, Domingue1991}. 
\citet{Buratti1995} drew similar conclusions for Ganymede and Callisto, suggesting their surfaces consist of a 80\% and 90\% void space, respectively.
%Modeling XX data, \citet{Belgacem2020} concluded that both hemispheres are consistent with ``\textit{a fine-grained, high-porosity regolith}''.
%the uniformity of texture over both hemispheres.
\citet{Hendrix2005} using ultra-violet phase curve analysis of Europa found a global porosity of $95\%$, with both hemispheres showing a similar high porosity, consistent with \citet{Domingue1997}.

\paragraph{Spectroscopy}
Spectroscopy measures the reflected or emitted light of a planetary surface across a range of wavelengths.
Spectroscopy in the near infrared probes depth up to 3 grains of the uppermost surface layer, which correspond to depth up to  $\sim 6 \, r_{\mathrm{g}}$. The retrieved parameters largely depend on the method used and their initial assumption as well as the quality of the data (and their associated uncertainties). A widely used approach is the linear unmixing, with which several compounds other than water ice were suggested \citep{Ligier2016, King_2022}. In such approach, each estimated spectrum is seen as a linear combination of different pure compounds at a particular grain size. While this method is suited to highlight the spectral contribution of particular compounds, it may lead to a wrong estimation of the surface properties by not accounting for the non-linear light-matter interactions occurring between the grains of the medium. Also, such method does not integrate the effect of porosity on the spectra as the surface is considered made of distinct patches of various surface proportions. Nonetheless, \citet{Ligier2016} stated that most of Europa's spectra can be fitted using grain sizes (\textit{i.e.} $2 \, r_{\mathrm{g}}$) ranging from $\SIrange{25}{200}{\micro \meter}$ and \citet{King_2022} found grain size in the $\SIrange{100}{1000}{\micro \meter}$ range. Both are compatible with our results.

To integrate the effect of porosity on the near-infrared spectra, more advanced radiative transfer models are needed, such as the Hapke model \citep{Hapke1993,Hapke2012a}. Using such approach, \citet{Mishra2021} have retrieved porosity values of 81, 85 and 97\% on three different spectra from the trailing hemisphere of Europa using Bayesian inversion. Such results were produced by considering very few compounds (only crystalline and amorphous water ice, and sulphuric acid octahydrate), thus highlighting the effect of a high porosity as a darkening agent, preventing the need for highly hydrated compounds.
%as the ones used in \citeauthor{Ligier2016} and \citeauthor{King2022}.

Similarly, results from \citet{CruzMermy2023} using the Hapke model and considering a porosity of 50\% have shown that only 4 compounds are required to fit one of the darkest spectra of Europa’s trailing hemisphere, and that many different combinations of salts can produce a similar fit to the data as long as water and sulphuric acid hydrates are included. Later, they showed that the grain size of water ice remains the same in all 173 different compound combinations that produce the same fit \citep{CruzMermy2025}, with a grain size (\textit{i.e.} $2 \, r_{\mathrm{g}}$) of about 500 $\mu$m for the crystalline water ice and 400 $\mu$m for the amorphous water ice. Such results are in agreement with the grain size estimated in \citeauthor{Mishra2021}, with a range of grain size of 100 to 900 $\mu$m for the amorphous and crystalline water ice, for different spectra of Galileo Near Infrared Mapping Spectrometer (NIMS) taken on the trailing hemisphere, without using any hydrated salts. Both these results are coherent with our grain radii estimate, and show how accounting for the porosity can lead to a very different estimate of the surface composition: a high porosity can be responsible for the darkening observed in the NIR spectra, and reduce the number of contaminants needed to fit the data.

\paragraph{Thermal and radar inversions}
\citet{Black2001b} have modeled Europa, Ganymede and Callisto radar reflectivities using different porosities: 25\%, 50\% and 75\%. While not significant, they have found "higher porosities [\textit{i.e.} 75\%] are slightly favored by the models for Ganymede and Callisto".
\citet{Travis2015} based on long-term thermal simulations of Enceladus, have included a "high porosity ice layer" on the surface that would provide a warmer near-surface environment, "in agreement with crater relaxation studies."
\citet{Lange2026} recently revisited Galileo PPR dataset, and also showed that some equatorial regions require porosities $>80\%$. %(\PPR data probes a bit deeper because it's not from eclipse analysis but still})
Based on the combined analysis of radar and microwave radiometry observations, \citet{Raza2026} results indicate that the south-polar terrains of Enceladus are covered by a highly porous water ice regolith, in agreement with plume fallout modeling \citep{Martin2023}. 
They inferred a regolith porosity ranging from $60\%$ to $90\%$ and grain radii $>\SI{500}{\micro m}$.

\paragraph{Others}
\citet{Nelson2018} notice the strong resemblance of the polarization phase curves of Europa with their samples exceeding 90\% porosities, suggesting that the surface of Europa may have a similar high porosity.
%\citet{Gundlach2018} mention that "Saturn’s icy moon Enceladus produces micrometre-sized water- ice particles by its cryo-volcanism."
Measurement of water ice grain size in the plumes of Enceladus showed ice grains in the $0.1-\SI{10}{\mu m}$ range \citep{Hedman2009, Ershova2024}.
Analysis of Cassini-VIMS measurements by \citet{Jaumann2008} showed that Enceladus particle size distribution is in the range $15-\SI{60}{\mu m}$ with a peak at about $\SI{20}{\mu m}$. \\

%Poch et al. (2018) performed polarimetric studies of ice particles, with their analysis showing that Europa most likely has sintered grains, leading to a more compact or less porous regolith. However, they caution that, unlike their pure water-ice lab samples, Europa’s surface contains a signiﬁcant amount of dark non-icy materials, which can signiﬁcantly affect the degree of polarization.

All of these different methods converge toward the same interpretation than our results: \textit{i.e.} an icy regolith of very high porosities ($>80\%$) and small grain radii ($<\SI{1}{mm}$).

\section{Discussion}
\subsection{What does the outermost regolith of icy moons look like?}

\subsubsection{The loosest sphere packing problem}

While remote sensing seems to agree that the upper centimeters of most icy moons is made of an extremely porous regolith, it is still unclear whether such porosities are realistic or even mathematically possible. 
Inverted parameters from remote sensing techniques are not guaranteed to be realistic, as assumptions and simplifications in the inversion methods can easily lead to non-physical results.

Indeed, \citet{Ferrari2016} stated that  ``\textit{As far as arrangements of spherical grains are concerned, [...] the upper limit of the porosity is limited to $\phi = 84\%$ for which the coordination number $N_{\mathrm{c}}=2$.}''
The restriction makes sense: to form a continuous structure, spheres must at least have two neighbors on average. 
However, we argue that the porosity limit at which this happens is not $\phi = 84\%$. 
This value, derived by \citet{Ferrari2016}, could come from a coordination number formula similar to the one found in SNOWPACK \citep{Lehning2002b} which gives $N_{\mathrm{c}}(84\%) = 2$. 
Yet this formula relies on empirical data from snow samples, which may vary between studies.
If we were to use, for example, \citet{Gusarov2003}'s formula instead (Table~\ref{tab:sphis}), we would find that even as porosity reaches 100\%, the coordination number never falls below 2: \textit{i.e.} $N_{\mathrm{c}}(\phi \to 1) > 2$, as shown in Appendix Figure~\ref{fig:coordnbr}.
However, would it make any sense for a material approaching $\phi\to 100\%$ to still maintain a non-zero coordination number? 

The answer lies in the field of mathematics.
The problem of finding the densest three-dimensional sphere packing has been long studied by mathematicians, notably since Johannes Kepler's conjecture in the 17th century, which established the lower porosity limit at $\phi = 25.95\%$ \footnote{The formal proof of the Kepler conjecture was only achieved four century later, in 2017 by \citet{HALES2017}}. 
Less famous, but more relevant here, is the problem of finding the \textit{loosest sphere packing}.
The solutions depend on the restrictions imposed, as we will see that without constraints, sphere packings of arbitrarily low density can be constructed.

\citet{Fischer2005} has shown that if the only restrictions are for spheres to be equal in size and number of contacts, it is possible to derive sphere packings where density approaches zero ($\phi\to 100\%$).

A stricter restriction would be to find the loosest mechanically stable sphere packing.
This problem has been nicely introduced by \citet{Dennis2022}: ``\textit{When sand is densely packed, it is strong enough to support the weight of an elephant. But how loosely can one pack sand before this rigidity is lost?}''
For a structure to be rigid, each sphere must touch at least four others ($N_{\mathrm{c}} = 4$), and the contact points must not all lie on one hemisphere or equator \citep{Gardner1966}.
For spheres of non-uniform sizes, \citet{Dennis2022} provided the solution: ``\textit{The answer is as loosely as one would like. That is, it is possible to rigidly pack hard spheres at any density, from filling all of space to filling none.}''  Their article present example of such structures built using bridges of spheres of arbitrary length that lead to asymptotically zero density.

\begin{figure}[htbp]
    \centering
    \includegraphics[width=\linewidth]{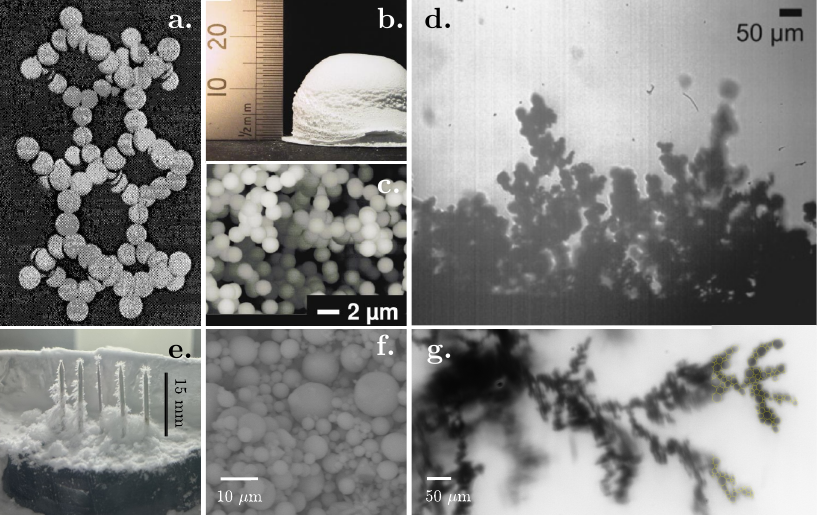}
    \caption{Examples of materials with extremely high porosities ($\phi >80\%$).  a) \citet{Heesch1933} $94.45\%$ rigid packing structure. b) and c) pictures from \citet{Blum2004} experiments of \ce{SiO2} agglomerates with $85\%$ porosity. d) Water ice aggregates with $89 \%$ porosity observed with a long-distance microscope \citep{Gundlach2011grains}.
    \textit{Bottom Row}: experimental pictures from the Core-Mantle Particle Sedimentation System  e) Cryogenically cooled chamber with $\SI{15}{mm}$ high needles that serve as sample mounts for ice aggregates.
    f) Micrometer-sized water ice spheres as seen from the scanning electron microscope. 
    g) Fractal ice aggregates as seen from the long distance microscope. 
    Each individual grain has been highlighted in yellow on the right side of the picture.
    }
    \label{fig:porous_examples}
\end{figure}

If we add the restriction that spheres must have a uniform size (as it is usually simplified to model the icy moons' regolith) and still want to search for mechanically stable structure, as of today, no structure of zero density has been found. However that does not mean that very high porosities can not be reached.
In 1932, \citet{Hilbert1932} described as what they believed to be the loosest packing, with a density of $87.7\%$.
A year later, this was surpassed by \citet{Heesch1933}, who detailed a much looser packing with a porosity of up to $94.45\%$, which remains as of today, the loosest packing respecting these constraints (see Figure \ref{fig:porous_examples}, a).

So mathematicians have long explored ways to pack spheres in three dimensions, and concluded that if the imposed restrictions are not strict enough, sphere packings of arbitrarily high porosity can be constructed. 
Even when restricting the search to rigid structures with uniform spheres, porosities exceeding 90\% are still possible. 
While such structure can be obtained mathematically, they are obviously artificially imagined, especially these arbitrary low density packings that require careful planning and construction tricks to be built.
Can high porosities sphere packings be found in nature, or is there a porosity threshold beyond which a natural packing of grains cannot exist?

%Random close packing of spheres in three dimensions gives packing densities in the range 0.06 to 0.65 (Jaeger and Nagel 1992, Torquato et al. 2000).

\subsubsection{Real life examples}

\paragraph{Terrestrial analogs}
On Earth, extremely high porosities are not rare for freshly fallen snow.
\citet{Fu2018} measured the average porosity of the snowpack in Northeast China ranging from 80\% at the beginning of the accumulation period to 70\% at the end of the accumulation period.
\citet{Clifton2007} reported values in the range 83\%–85\% for fresh wind-blown snow.
\citet{Colombo2023} mapped thermal inertia to snow density in the Western European Alps.
The authors found that fresh snow can have thermal inertia as low as $\SI{80}{}$ and measured density as low as $\SI{100}{kg.m^{-3}}$, which based on Equation \ref{eq:density}, relates to a porosity of $89\%$.

\paragraph{Experimental analogs}
Experimental planetology has also produced icy moons analogs of extremely high porosities with arrangement of spheres.
\citet{Blum2004} using monodisperse \ce{SiO2} sphere in the micrometer size have produced highly porous agglomerates by ballistic \textit{hit-and-stick} deposition (see Figure \ref{fig:porous_examples}, b and c).
Because of the low impact velocities during sample preparation and the large interparticle adhesion force, the agglomerates were able to reach a porosity of $85\%$ while being mechanically stable against unidirectional compression.
Following the same method, \citet{Krause2011} measured the thermal conductivity of very porous and fragile dust samples (up to $85\%$ porosity).
\citet{Schrapler2021} produce monodisperse highly porous water-ice by sedimentation of water-ice particles.  With their experimental technic they were able to form an agglomerate with a mean porosity reaching up to $90\%$.
Under Earth gravity, \citet{Gundlach2011grains} performed a sedimentation experiment, of micrometer-sized water ice onto a target which produced 89\% porosity samples due to what they also described as a \textit{``hit-and-stick''} behaviour (see Figure~\ref{fig:porous_examples} d). Without Earth's gravitational restructuring, \textit{i.e.}, at icy moons' lower gravity, we would assume even higher porosities.
\citet{Nelson2018} using powdered aluminum oxide, manage to produce monodisperse samples reaching porosities as high as $97.7\%$. 
They explain that porosity increases inversely proportional to the grain size in their experiments because of the lower gravitational forces that scale as $r_{\mathrm{g}}^3$.
% repulsive vdw??

To complete these studies, here, we also present ongoing experimental work using a Core-Mantle Particle Sedimentation System (CoMPaSS) (see Figure~\ref{fig:porous_examples}, \textit{bottom row})
The CoMPaSS experiment is designed to measure the surface energy of micrometer-sized water ice and core-mantle particles under cryogenic conditions ($\SIrange{90}{200}{K}$). By linking microscopic adhesion with macroscopic aggregate behavior these experiments will help understand planetesimal formation.
In the experiment, micrometer-sized water droplets are produced using a disperser and injected into a cryogenically cooled chamber, where they freeze almost instantaneously. The resulting micrometer-sized ice particles sediment toward the bottom and stick to a needle (tip \SI{2.5}{\meter} and height \SI{15}{\milli\meter}) that serves as a sample mount (see Figure~\ref{fig:porous_examples}, e, f). There, they form fractal ice aggregates which are imaged using a high-resolution camera in combination with a long-distance microscope (LDM) (see Figure~\ref{fig:porous_examples}, g).
Although it is not the primary goal of the CoMPaSS experiments, this preliminary imaging reveals such high porosity aggregates in high resolution.
These long fractal structures, surrounded by large voids, occur due to the high cohesivity of water ice at these temperatures, allowing porosities as high as those reported in \citet{Gundlach2011grains} ($\sim 89\%$).

\paragraph{In situ planetary ice}
As of today, the only \textit{in situ} interaction with planetary ice was made at the surface of comets.
The Philae lander on Rosetta revealed extremely porous and mechanically weak ice. Analyzing the lander's impact on the regolith, \citep{ORourke2020} estimated the ice porosity of comet 67P/Churyumov-Gerasimenko to range between 68\% and 82\%. Other independent methods using the Philae Consert radar \citep{Herique2019}, the Rosetta Radio Science Instrument \citep{Paetzold2016} and a planetesimal model \citep{Fulle2016} also estimated the porosity of the comet nucleus to be particularly high, ranging between 63\% and 85\%.
We note that using the same model as the CONSERT team, \citet{Lethuillier2016} found a surface porosity $<50\%$ for the comet, more compacted than its interior, likely due to micrometeorite impacts.

\paragraph{What could explain such extreme porosities?}
Considering the small particles composing the regolith, \citet{Wood2020} explained that, under the low gravity of icy moons, cohesive surface forces dominate over gravitational forces. 
Monte Carlo simulations of monodisperse ice packing \citep{Yang2000} have revealed that strong adhesion allows particles to form open, chain-like structures surrounded by large pore spaces.
This hypothesis aligns with work by \citet{Jabaud2023}, who estimated the Bond number (ratio of adhesive to gravitational forces) on icy moons. Although temperature-dependent due to surface energy variations, the Bond number is very high, implying that fresh deposits on Europa behave as highly cohesive structures, unaffected by the negligible gravitational force but that would be easily disrupted by external forces.
Based on these results we conclude that because ice on icy moons is likely very cohesive it allows it to reach extreme porosities.
Still, the structure is probably not very rigid, analogous to the regolith on comet 67P/Churyumov-Gerasimenko  \citep{ORourke2020}.

% SECTION LAYERING
\subsection{Surface layering, regolith compaction and formation scenarios}
\label{sec:layering}

%The electrical skin depth is the distance at which the power of an electromagnetic wave (of a given wavelength) is attenuated by a factor 1/e ≈ 37%. Beyond this distance, the contribution of deeper layers to the measured signal becomes
Since decades, space instruments operating at different frequencies have observed the Galilean moons at various depths.
For example, on Europa, the day length ($\sim 80$ hours) results on a thermal skin depth in the order of a few to dozens of centimeters depending on the thermal properties of the surface.
However, for short phenomena such as eclipses, the period is much shorter, usually a few hours (\textit{e.g.}, Europa’s eclipse duration is $\sim 2,8$ hours). 
This results in a significantly shallower skin depth, which is why eclipses observations in the 1970s probed only the first few millimeters of the Galilean moons (see Table~\ref{tab:lowTI}).
These eclipse observations revealed a thermal inertia close to $\Gamma_{\mathrm{LTIM}} = \SI{15}{}$.
While observations made during a diurnal cycle are typically in the order of $\Gamma \sim 50-100$ \citep{Spencer1999, Rathbun2010, Lange2026}, and ALMA microwave observations lead generally to larger thermal inertia $\Gamma > 100s$ \citep{Kleer2021, Camarca2023, Thelen2024}.
%I need to double check the exact numbers. Update it's 56-184 from Thelen2024 on Europa}). 

\begin{figure}[!ht]
    \centering
	\includegraphics[width=0.72\linewidth]{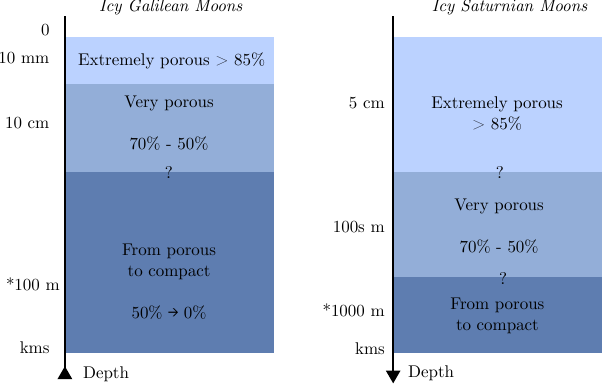}
    \caption{Potential vertical layering of the regolith as suggested by the estimated thermal inertia in the literature (see Table~\ref{tab:lowTI}). The uppermost, extremely porous layer has been estimated from eclipse observations on the icy Galilean moons. The exact transition from porous to compact ice is not known, but base on self-weight compaction models under gravity \citep{Mergny2024c} it is not expected to happen before the first hundreds of meters, well below the near-surface probed by remote sensing.}
    \label{fig:porous_layering}
\end{figure}

While it is not clear where this transition occurs and whether it is smooth or discontinuous, we observe a clear increase of thermal inertia between the first millimeters and the first centimeters of the regolith.
If we use the thermal properties models used in our study along with our assumptions, this implies a transition from extremely high porosities ($>80\%$) in the top millimeters to moderately porous layers ($70\%-50\%$) at centimeter depths \citep{Thelen2024}. 
An illustration of such layering is presented in Figure~\ref{fig:porous_layering}, Left.
Even, regardless of agreeing with the conductivity models presented in the present article, simply observing this gradient in thermal inertia with depth indicates the presence of layering within the first millimeters to centimeters of the Galilean moons' regolith.
This increase of thermal inertia with depth presents notable paradoxes:
\begin{itemize}
  \item  Such compaction happens at deeper scale for Saturnian moons, not centimeters but meters scale \citep{LeGall2014, Bonnefoy2020}). Their highly porous regolith seem to extend to substantially greater depths (see Figure \ref{fig:porous_layering}, Right).
	\item  At such small scales, gravity has absolutely no effect on compaction: \citet{Mergny2024c} demonstrated that self-weight compaction of Galilean moons regolith may only becomes noticeable at depths exceeding $\sim\SI{100}{m}$.
	\item The uppermost layer of the regolith is the most exposed to solar heating: maximum temperatures are reached at the surface and decrease with depth exponentially following the thermal skin depth. As a result, ice sintering which increases thermal conductivity should be more efficient at the surface than at depth as modeled by \citet{Mergny2024i}.
  \item The uppermost layer of the regolith is also the most exposed to charged particles and micrometeorites bombardment, which according to some studies may compact the near surface \citep{Raut2008}. 
\end{itemize}
These points lead to an opposite conclusion: if compaction were occurring, it should follow the opposite trend: thermal inertia should be the highest at the surface and decrease with depth.
Then, why do space observations suggest the presence of a more insulating layer at the millimeters depths?
Which physical processes could be responsible for such small-scale compaction of the regolith ? 
Here we explore three potential formation scenarios of the Galilean icy moons regolith.

%Mishra2021 Since reﬂectance spectroscopy at the Galileo NIMS wavelengths is probing just the upper hundreds of microns to millimeter of the regolith, it is reasonable to think that the surface we are seeing is highly porous or “ﬂuffy” in a thin layer at the top, but with compactness varying with depth and perhaps increasing.
%Nelson: The first possibility is that the uppermost surface layer, with a depth on the order of the wavelength, may have a high porosity such that the vacuum–surface interface is not a sharp discontinuity in refractive index. The other possibility is that the surface is very rough and cannot be described as “smooth” at the wave- length scale.
%Martin: On an ice-rich body like Ceres, the process of impact gardening has been found to be an order of magnitude slower than on the Moon, suggesting that Ceres does not have meters of impact-generated regolith like the Moon (Costello et al., 2021).

%Howett2022: this is different from the Jovian system, where eclipse-derived thermal inertias are much lower than those derived from diurnal studies. The cause of this difference is not known, but one possibility is that the E-ring grains that bombard Dione’s leading hemisphere overturn it,

\begin{figure}[!ht]
    \centering
    \includegraphics[width=\linewidth]{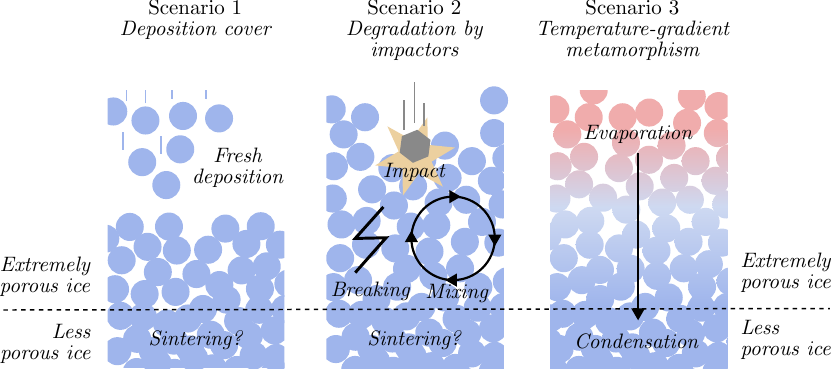}
    \caption{Three potential formation scenarios for the regolith to explain the shallow layering of the Galilean icy moons. These scenarios are not mutually exclusive, and the actual processes may involve a combination of one, two, or all three.}
    \label{fig:scenarios}
\end{figure}

\subsubsection{Deposition Cover}

In this scenario, an older, compacted layer is progressively covered by a fresh fallout of new regolith:
\begin{itemize}
    \item The initial condition is a uniform but extremely porous regolith.
    \item Isothermal sintering occurs, increasing the grain contact areas and thus the thermal inertia of the regolith. The regolith becomes more thermally conductive (its density does not change, which is a common misunderstanding of the isothermal sintering process).
    \item A fresh fallout of new, extremely porous ice (e.g., from plumes or the continuous flux of micrometeorites ejecta) then covers the sintered regolith, creating the layering.
    \item This cycle repeats, with the competition between sintering and fallout rates determining the thickness of the uppermost layer (millimeter to centimeter range, see Figure~\ref{fig:scenarios}, left).
\end{itemize}
For this scenario to work, isothermal sintering must neither be much slower nor much faster than the fallout rate, otherwise no layering would be observed. 
%(since the estimated thermal inertia of this layer is incompatible with sintering).
Since sintering is highly temperature-dependent (and thus latitude- and albedo-dependent), the layering should vary regionally, affecting the transition depth (see Figure~\ref{fig:scenarios}, middle). 
This would also explain why such layering is absent on the Saturnian moons: temperatures are too low for sintering to occur.
Although some regions, like Enceladus Tiger Stripes, are warm enough for sintering to be efficient, plume fallout rates and deposition of E-ring particles on the trailing side of Mimas and leading side of Tethys, Dione and Rhea, might be faster \citep{Gundlach2018}.

%However, the lack of significant differences between the Galilean moons makes this scenario unlikely. 
%
%"Micrometeoroid impacts combined with the sputtering by solar wind particles are a significant source of regolith grain loss (Sasaki et al., 2001), particularly on bodies with little or no atmosphere (Yamamoto, 2002).
%" Meteroid and dust bombardment causes the regolith particles to eject at different velocities: low ejection velocities lead to particles falling back to the surface and redistributing grains in the regolith "

%This is supported by:
%Europa regolith are not dominated by meteoritic gardening on centimeter scales, but proba- bly are depositional. The porosity, and therefore grain size distribution, is uniform over the entire surface.
 
%Gundlach2018
%Additionally, the active cryo-volcanism leads to a renewal of the surface, which would require an even faster sinter process. Thus, we conclude that Enceladus possesses a powder-like surface consisting of unsintered micrometre-sized par- ticles, which was also indicated by an earlier work that compared the moon’s phase curve with a granular ice sample in the laboratory (Jost et al. 2013).

%Ferrari2018:Also I'll cite Ferrari2018 "Tit can be easily reproduced if heat trans- fer is dominated by radiation in pores, despite low temperatures, because the conduction through grains is limited, either due to the presence of amorphous ice or because of the roughness of grains."

\subsubsection{Degradation by impactors}

In this scenario, a compact[ing] regolith is eroded by the constant flux of impactors hitting the surface. A first simple approach to consider would be:
\begin{itemize}
    \item The initial condition is a uniform material, whether porous or not. A compact material could be for example one produced by cryovolcanism.
    \item This initial surface is broken down into a regolith by the ongoing flux of micrometeorites, which fracture the ice, mix and overturn the material, and eject particles that redeposit on the surface \footnote{if redepostion from ejecta is consequent then Scenario 1 is also involved}.
    \item The degradation of the initial structure due to meteorites scales with their penetration depth, thereby creating a layering.
\end{itemize}
Micrometeorite bombardment is widely considered as the primary process responsible for the regolith formation on the Moon and Mercury, so it is reasonable to assume it also the case on the Galilean icy moons. 
However, \citet{Domingue1991} noted that if the Galilean moons' regolith were only the result of micrometeorite bombardment, we would expect porosities similar to those of the lunar regolith, which are around 60\%. 
Yet, here we have previously gone through multiple evidence which indicate that the regolith of icy moons is likely much more porous ($>80\%$).
Additionally, if micrometeorites were the only responsible for layering, we would expect the transition of thermal inertia to occur at greater depths than the first few millimeters. 
For example, micrometeorites change the thermal conductivity of the lunar regolith up to depths of 10 cm by a factor of 5 to 10 \citep{Wood2020}.
These inconsistencies suggest that another process must be involved to reduce the transition depth. Let us consider then the updated scenario:
\begin{itemize}
    \item The initial condition remains the same: a uniform material, whether porous or not.
    \item The surface is broken down into a regolith by meteorites bombardment.
    \item A mechanism, such as isothermal sintering, increases the overall conductivity of the regolith \citep{Gundlach2018}, with efficiency following the thermal skin depth.
    \item The cycle returns to step 2: Erosion occurs either due to micrometeorite impacts, which break the sintered bonds and redistribute the material \citep{Costello2021}, or through the flux of charged particles that erode the surface, with re-deposition occurring via Scenario 1.
\end{itemize}
The competition between erosion and a conductivity-increasing mechanism like sintering results in a layered structure: an untouched deeper sintered layer beneath a mixed and broken upper layer (see Figure~\ref{fig:scenarios}, middle). 
This layering would depend on the micrometeorite flux, the sintering rate, and, if applicable, to the erosion rate from charged particles. Thus, it is region-dependent.

\subsubsection{Temperature gradient metamorphism}

Sintering is a term that encompasses a wide range of processes altering the microstructure of ice grains. It results in the growth of the contact area between grains, \textit{i.e.}, the growth of the bond. The responsible physical processes are diffusion mechanisms, either from the bulk or the surface of the grains.
The only sintering process studied in planetary science, whether analytically \citep{Molaro2019}, numerically \citep{Mergny2024i}, or experimentally \citep{Molaro2019, Gundlach2018, Choukroun2020}, is isothermal sintering under closed-pore conditions via evaporation-condensation (referred to as sintering here). To date, all focus in the planetary science field has been on this single process. The reason for this is that evaporation-condensation is known to be the dominant diffusion mechanism initiating bond growth on icy moons. However, all current models assume a closed system where grains sublimate and material recondenses only on neighboring bonds. Yet sintering on Earth only occurs in snowpacks where pores are isolated (closed-pore structures).

In porous snow, if porosity is high enough, pores can communicate and vapor can be transported between them. Since the surface of icy moons seems to be very porous, this suggest that the regolith has such open-pore network.
As \citet{Mergny2024i, MergnyThesis2024} suggested, if pores can communicate, a diffusion process known on Earth as \textit{temperature gradient metamorphism} (hereafter shorten to \textit{TG-metamorphism}) or also called kinetic growth metamorphism could occur. Unlike isothermal sintering, this process involves vapor sublimating from warmer layers and condensing on colder ones. This would also produce latent heat that could affect the thermal balance of the regolith.

On Earth, temperature gradient metamorphism becomes significant when gradients exceed $\SI{10}{K.m^{-1}}$ \citep{Colbeck1983}. On Europa, \citet{Mergny2024h} showed that temperature fluctuations can reach up to $\SI{15}{K}$ within the first $\SI{10}{cm}$, well above the Earth-based threshold for TG-metamorphism. Temperature differences create water vapor concentration gradients within the pore space. If the pore space is large enough, a substantial vapor flux develops from warmer to colder regions. 
We thus consider the following scenario:
\begin{itemize}
    \item The initial condition is a uniform regolith with an open-pore structure.
    \item  TG-metamorphism occurs: grains in the uppermost layer sublimate vapor into the pore space. Due to the temperature gradient, vapor pressure is lower in deeper, colder layers, creating a pressure gradient that transports material from the uppermost layer to deeper layers. Condensation in these colder layers increases thermal inertia at depth (see Figure~\ref{fig:scenarios}, right).
\end{itemize}
Recent results from \citet{Lange2026} showing latitudinal variations in thermal inertia from PPR analysis, could be explained by TG-metamorphism being more efficient at low latitudes. 
If TG-metamorphism can be a defining process of the regolith, this highlights the need for more experimental data and modeling studies to estimate its effect on icy moons.

If vapor is unable to diffuse vertically either due to the very low pressure involved or due to tortuosity, TG-metamorphism might not be happening. If TG-metamorphism is somewhat impossible but sintering is, an alternative scenario, that does not involve exchange of matter between layers could also induce surface layering:
\begin{itemize}
    \item The initial condition is a uniform regolith with high enough porosity.
    \item Isothermal sintering induces sublimation everywhere. Grains in the uppermost layer sublimate vapor into the tenuous exosphere, decreasing density and as a result increasing porosity of the uppermost layer. Meanwhile, grains from deeper layers also sublimate but the vapor re-condenses locally. This process, which we will name \textit{vacuum-based sintering}, has been observed experimentally by \citet{Gundlach2018}.
\end{itemize}

\subsubsection{A complex interplay of surface physics}
Through the three scenarios presented here, we have explored how different surface processes could explain the small-scale compaction observed on the Galilean moons.  
%This may not be an exhaustive list as other surface processes not described here may also contribute to the regolith's layering.
While these scenarios were presented independently, they are not mutually exclusive.
In reality, the regolith may be subject to all three simultaneously, with varying degrees of efficiency. 
Ice metamorphism, whether under isothermal or temperature-gradient conditions, is likely to occur due to the high temperatures found. According to \citet{Mergny2024i} when daily temperature reach $\SI{115}{K}$ sintering effects can be seen in only $\SI{1}{Myr}$.
At the same time, the ongoing flux of micrometeorites and charged particles erode the surface, break the sintered bonds and mix the material (Scenario 2). Then, new material originating either from plumes or more likely from ejecta, redeposit on the surface and cover older material (Scenario 1). 
The resulting picture is a complex coupling of these processes, which makes its modeling challenging. 

Finally, although the presented scenarios can explain the surface's layered structure, if interpreted under the assumption of a regolith thermally dominated by hexagonal water ice (Assumption~\ref{as:cry_dom}). However, additional scenarios emerge if impurities are considered, such as a vertically varying chemistry with insulating exogenic material concentrated at the uppermost surface and diminishing sharply within the first few centimeters. 
This highlights the need for more advanced multiphysics models, or improvements to existing ones like \textit{LunaIcy} \citep{Mergny2024i}, as the surface of these moons is shaped by a dynamic interplay of multiple surface physics.

\subsection{Implications for exospheres and interiors}
\subsubsection{Implications on volatile trapping and evolution: the case of molecular oxygen}
%Based on our results, clearly the surfaces of icy satellites are far more porous than Earth's ice at $\sim$ meter depths. 
High porosity is advantageous for the creation of voids needed for trapping volatiles, as appears to be the case for molecular oxygen, O$_2$ dimers, at Europa and Ganymede \citep[their figure 3]{Oza2025}.
Indeed, the recent volatile trapping model suggests that since volatiles are actively destroyed by energetic charged particles in Jupiter and Saturn's magnetospheres \citep{Johnson1990} with their steady state abundance being regulated more so by how fast ice crystallizes and traps them. 
Trapping is enhanced when porosity increases, which also led to the inference that Europa at least, must have molecular oxygen trapped at depth \citep{Oza2018}. This is also likely the case at Ganymede \citep{Leblanc2017} and Callisto \citep{CarberryMogan2022}. Trapping of volatiles such as CO$_2$, must also be occurring at the Saturnian satellites Dione and \citep{Ahrens2022} and comets \citep{Oza2024}. 
In \citet{Oza2025} the bubble radius, which constrains the trapped volatiles, is roughly constrained to be $\sim \SI{10}{nm}$ roughly in line with the contact radius for a $\mu$m sized grain at 125K with a water surface tension of 0.08 J m$^{-2}$ close to the canonical value of 0.076 (red line, Figure 3). Consequently, if the water surface tension varies thermally, this value may affect the bubble radius by up to two orders of magnitude at 100 K \citep{Jabaud2023}.

\subsubsection{Regolith as a thermal bottleneck for the interior}

\begin{figure}[htbp]
    \centering
    \includegraphics[width=\textwidth]{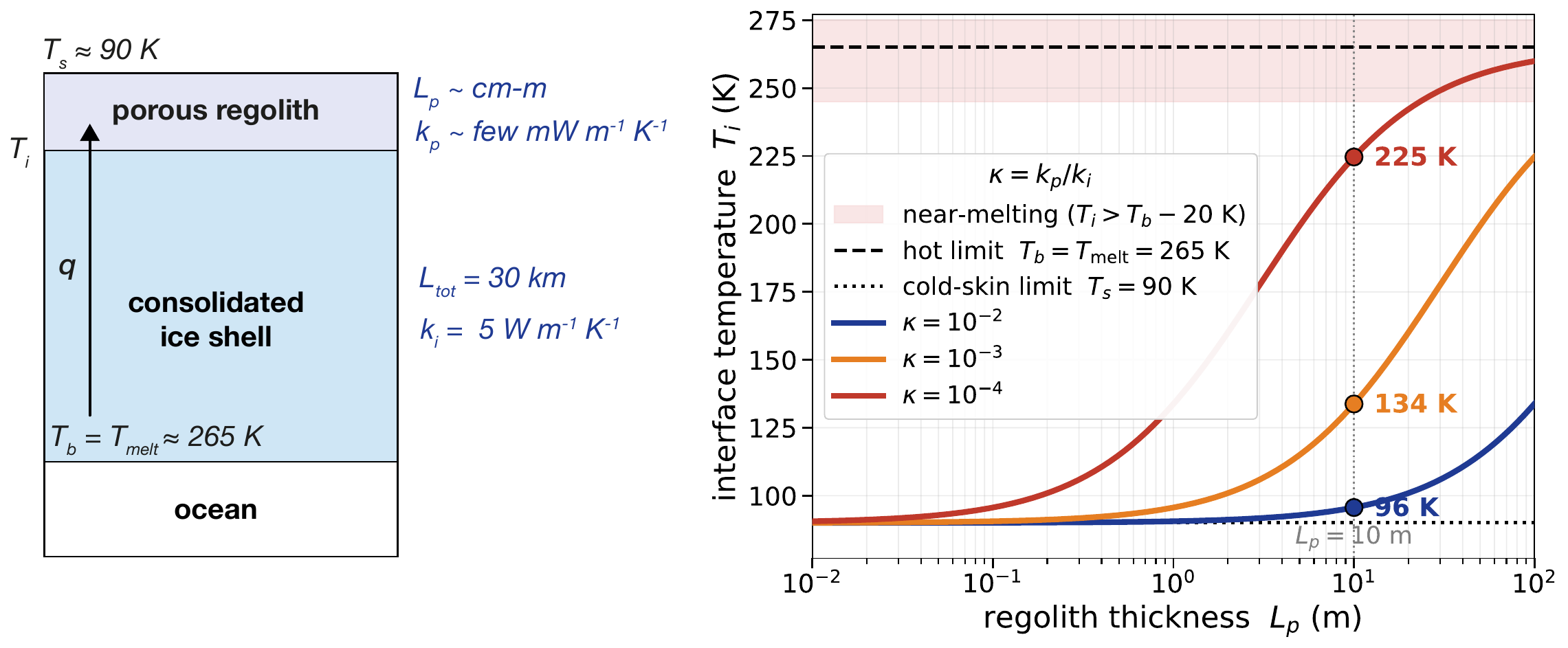}
    \caption{Thermal effect of an icy regolith layer on the temperature structure of an icy moon's conductive lid. Left: Schematic of the two-layer model. A thin porous layer sits on top of kilometers-thick consolidate ice layer, bounded by the surface and the ocean. The temperature at the regolith-ice interface, $T_i$, is the unknown set by the relative thermal resistances of the two layers. Right: Steady-state interface temperature, $T_i$, as a function of porous layer thickness, $L_p$, for three values of the conductivity ratio $\kappa$. As insulation increases (larger $L_p$ or lower $\kappa$) $T_i$ rises from the cold limit towards the melting point. The pink band marks the near-melting point regime. Filled circles indicate the reference case, $L_p=10$ m. Even a meter-scale regolith is able to push $T_i$ deep into the warm region if $\kappa\leq10^{-3}$.
    }
    \label{fig:regolith_interior}
\end{figure}

Although a potential icy regolith layer occupies a vanishingly small fraction of the total lid thickness on icy moons (centimeters to meters on top of a kilometers-thick outer ice shell), its low thermal conductivity (suggested by the low thermal inertia values, for details see Section~\ref{sec:methods}) makes it inevitable to consider for the thermal state of the lid. In steady-state two-layer conduction the surface temperature, $T_s$, is set externally and the basal temperature, $T_b$, is pinned to the pressure-dependent ice melting point. The conductive heat flux through the lid is then fixed by the thermal resistance $R=L/k$ (in m$^2$ K W$^{-1}$). What varies with regolith properties is the temperature at the interface between regolith and consolidated ice, $T_i=T_s+F(T_b-T_s)$, where the temperature drop fraction $F=R_p/(R_p+R_i)$ reduces to a single function of two dimensionless ratios: a geometric ratio $\lambda=L_p/L_{tot}$ and a conductivity ratio $\kappa=k_p/k_{i}$, where $L_p$, $L_{tot}$, $k_p$, and $k_i$ are the porous layer thickness, the total thickness, the porous layer conductivity, and the conductivity of bulk ice, respectively. For plausible icy-moon values ($k_p$ of a few mW m$^{-1}$ K$^{-1}$, $k_i\approx 5$ W m$^{-1}$ K$^{-1}$ at the surface, so $\kappa\sim10^{-3}$), even a meter-thick regolith pushes $T_i$ from near cold surface temperatures toward the melting point (see Figure~\ref{fig:regolith_interior}). It should be noted that this effect is amplified if the underlying ice shell convects in stagnant-lid mode: the conductive layer then comprises only the stagnant lid ($\sim10$ km, rather than the full shell), capped at its base by the temperature where the ice becomes warm enough to be ductile, which is a few tens of Kelvin below the melting temperature. With a thinner conductive layer, the regolith dominates the thermal resistance more strongly and $T_i$ rises even faster toward this near-melting base temperature of the lid. This has direct consequences for ice rheology, the depth at which convection can set in, and constraining the survival of liquids in the ice shell.

\subsection{Assumptions, simplifications, limits and future improvements}

\label{sec:limits}
\subsubsection{Assumptions and simplifications}

The only way to directly measure microstructural properties such as porosity and grain radius on icy moons would be through \textit{in-situ} analysis of the regolith. 
At the very best, such measurements are planned to happen in the late 2050s, when ESA’s L4 mission lander is expected to reach the surface of Enceladus. 
Until then, estimates of these properties rely indirectly on inversion models derived from remote sensing data, complemented by laboratory experiments on analogs and numerical simulations.
All these models inherently involve assumptions and simplifications (which may not always be explicitly stated), that introduce potential caveats in the estimation of surface properties.  Here we draw a comprehensive list of the hypotheses and simplifications made in our study, along with their impacts on our results.

When conducting such exercise, it is useful to classify each parameter based on how directly it relies on raw data versus how much it depends on multiple layers of modeling assumptions (similar to the distinction in computer science between low-level, close to the machine languages and high-level, human-readable languages).
For example, the rawest data in thermal analysis is the instrument counts or voltage. Still quite raw, is the thermal flux measured by the instrument, which assumes only that the instrument is well-calibrated and observing the correct region. At the extreme opposite is porosity, a derived parameter from numerical inversions which relies on multiple modeling assumptions.\\

% Assumption 1
The first large assumption that we have made in our study is that:
\begin{assumption}
   The low thermal-inertia of icy moons found in Table \ref{tab:lowTI} have been correctly estimated. 
   \label{as:correctTI}
\end{assumption}

Thermal inertia is not a measured property but a derived property relying on many assumptions made in the inversion models. 
This assumption implies that no systemic bias exists in the inversion models. Yet, there remains the possibility that all these models could share a common simplification, such as the widely used assumption that IR emissivity lies in the range $\epsilon \sim 0.9 – 1$. %(\textit{e.g}., vertical homogeneity or neglecting surface roughness, etc...).
% Mention the fact that models inverting thermal inertia only may have high degeneracy ?
Supporting this possibility, recent JWST Mid-Infrared Instrument (MIRI) observations of Ganymede \citep{BockeleeMorvan2024} encountered difficulties in thermal inversion: using measured flux and a Bond albedo map from \citep{Kleer2021}, a realistic thermal inertia could not be derived. Even after accounting for surface roughness, inverting realistic values required the authors to reduce Ganymede’s albedo by $\sim 35-40\%$. This may be due to issues with the albedo map or from a simplification like the infrared emissivity assumption.
%It is beyond the scope of this paper to review all potential simplifications made in studies deriving LTIM values. 
Here, we relied on the fact that these \textit{Low Thermal inertia on Icy Moons} (LTIM) have been estimated consistently across all icy satellites and using a variety of instruments and modeling techniques. 
Since the literature shows strong agreement on these values, this encourages us to believe that there are correct and serve as the basis of our analysis.
%this adds to often brutal assumptions on the IR emissivity and the Bond albedo which have a direct effect on the estimate of the thermal inertia.

Finally, we adopted the upper limit $\Gamma_{\mathrm{LTIM}} = \SI{15}{}$ in this study, in order to consider a common LTIM value for all icy moons.
This value is slightly higher than most estimates for icy moons (see Table~\ref{tab:lowTI}). This conservative choice demonstrates that even in this case, we obtain very tight constraints on the regolith microstructure.
For example, values such as $\Gamma = 12 \pm \SI{3}{}$ for Ganymede or $\Gamma = 10 \pm \SI{1}{}$ for Callisto would lead to even tighter constraints, requiring even higher porosities. \\

The second major assumption in our analysis is that:
\begin{assumption}
    The thermal properties of the regolith are dominated by crystalline water ice (\textit{i.e.}, thermal contributions from other compounds are neglected).
    \label{as:cry_dom}
\end{assumption}
This assumption is commonly adopted in the literature when deriving porosity \citep{Trumbo2018, Kleer2021, Camarca2023, Thelen2024, Lange2026}. 
While one reason for this assumption is the current lack of models and experimental data to correctly describe the thermal behavior of porous ice mixture, it is also grounded in some valid observations: 

First, the $\Gamma_{\mathrm{LTIM}}$ values were derived across the full hemisphere of each icy moon, meaning they average regional composition. Second, all icy moons show this very low thermal inertia, and share in common that their surface is primarily composed of crystalline water ice.
One might argue that a common compound (such as dust or amorphous ice) could be invoked instead of extremely high porosities to explain the low thermal inertia. 
However, such a compound should, either be present across all icy moons and their entire observed surfaces or at least, each moon would require a compound with similarly insulating properties.
For example, amorphous ice could be a good candidate: it significantly reduces the thermal conductivity of water ice and, where it exists, it likely influences the regolith thermal inertia, as demonstrated by \citet{Ferrari2016}.
But then, how to explain the low thermal inertia  of regions whose spectroscopy suggests a surface dominated by crystalline water ice?  

Icy moons are obviously not only composed of crystalline water ice, and crystalline water ice is not the only factor influencing conductivity. In this paper, we explored the implications of this assumption for the regolith structure. Then, we identified two arguments by contradiction that arise if this assumption is not made:
\begin{itemize}
    \item  First, the extremely high porosities we obtain are compatible with photometry and spectroscopy studies, as shown in Section \ref{sec:lit}. 
    If impurities were invoked as an insulating material, they would also reduce the derived porosities, making them incompatible with other literature approaches. The only way to reconcile these studies would be that these impurities also affect optical properties in the exact same way, which we consider unlikely. 
    \item  Second, \citet{Lange2026} pointed out that ``\textit{Studies focusing on the thermal conductivity of porous mixtures of ice and silicate minerals (representative of cometary materials) have shown that the presence of minerals within porous ice can drastically increase its thermal conductivity at low temperatures.}''  If impurities increased conductivity, even higher porosities and smaller grain radii would be required to satisfy $\Gamma < \Gamma_{\mathrm{LTIM}}$. Given that our model with pure crystalline water ice already approaches the upper porosity limit ($>90\%$) and minimum grain radii ($\sim 1 \mu m$), we conclude that conductivity-increasing impurities would lead to unrealistic values of porosities and grain radii.
\end{itemize}

Nevertheless, we acknowledge that darker surfaces are less likely to be composed of pure crystalline water ice, limiting the relevance of this study for such cases. For example, \citet{LeGall2024} showed that a dust layer over water ice can reproduce Iapetus' temperature curve, suggesting our assumption may not apply there.\\

%The second argument is that  \citet{CruzMermy2023} Bayesian inversion showed that water ice has much larger grain sizes than other compounds. 

The third major assumption in our analysis is:
\begin{assumption}
    The models for radiative and contact thermal conductivity are correct and valid under the temperature and environmental conditions of icy moons.
    \label{as:correct_model}
\end{assumption}

These models rely on various simplifications, such as perfectly spherical grains with uniform sizes.
\citet{Piqueux2009a} pointed out that natural particles should have surface irregularities that will increase the contact areas between grains. While we did not include such effect here, this would increase the contact conductivity and thus give even better constrains on the regolith microstructure.
The analytical contact models that we have used have also been validated using analog experiments with \ce{SiO2} \citep{Krause2011} grains rather than \ce{H2O}, and typically at much lower porosities (for example, \citet{Gundlach2012} conductivity model has been validated with \ce{SiO2} samples of 30-40\% porosity). Quantifying the impact of these simplifications is challenging and would require further comparisons with icy moon analog experiments and refined models.

We note that the constraints on the maximum grain radius shown in Figure~\ref{fig:uppergrain} do not depend on any contact conductivity model. Thus, they provide a lower-level estimate of the maximum grain radii, requiring fewer assumptions.

While each contact conductivity model introduces different thermal conductivity values, Figure~\ref{fig:all_TIs} demonstrates good agreement among models in the high-porosity range. 
We therefore made an arbitrary choice of contact conductivity model with Gundlach2012 model representing one of the least conductive option in the porosity range of interest ($\phi \in [0.7, 0.9]$). Choosing a more conductive model, such as \citet{Adams1993} or \citet{Shoshany2002}, would only impose even better constraints, again requiring higher porosities.
% A general trend in this paper is to always consider the most conservative case.

\subsubsection{Limits and perspectives}

Due to the assumptions previously mentioned and the simplifications of the current thermal conductivity models, our results could be improved in a few aspects:

\paragraph{Gas conduction}
Here, we assumed that gas conduction is negligible, as often invoked in the literature, given the very low surface pressure on icy moons.
Sublimation and recondensation, likely to happen on the surface of the hottest icy moons, will produce latent heat that will be thermally transferred in gas phase.
In cases where radiative and contact conductivity are also extremely low, gas conduction might thus not be negligible. It would be interesting to model its contribution and explore the constraints imposed by $\Gamma < \Gamma_{\mathrm{LTIM}}$ in this scenario. This could allow us to estimate upper limits for gas pressure and diffusion within the pores.

\paragraph{Surface Tension}
The surface tension greatly influences regolith conductivity, as it defines the quality of contact through grain radii.
Since we used new data from \citet{Jabaud2023}, which is multiple orders of magnitude lower than previously assumed \citep{Ferrari2016, Molaro2019}, this reduces the bond radius by several orders of magnitude as well.
If we were to use $\gamma \approx \SI{0.06}{J.m^{-2}}$ instead, we would obtain a higher contact quality, and thus, for the same grain radius, temperature, and porosity, a much higher thermal inertia. This would result in much better constraints, requiring even higher porosities and smaller grain radii. Following our study’s general trend of choosing the most conservative case (here, Jabaud’s expression), we show that this still allows us to place strong constraints on the regolith.

We note, however, that our constraints are very relaxed at low temperatures (Figure~\ref{fig:param_allowed}, at 50 K) because \citet{Jabaud2023} surface tension predicts extremely small contact areas. In these cases, the bond radius can become smaller than $\SI{1}{nm}$, approaching the size of a single water molecule ($\sim\SI{0.28}{nm}$). We consider these bond radii to be unrealistically small.
This suggests either that the JKR model \citep{Johnson1971} is not valid under these conditions or that surface tension is actually higher. In either case, if bond radii at low temperatures are underestimated, thermal inertia is likely underestimated as well, meaning that, in reality, we would have even tighter constraints on porosity and grain radii at 50 K.

\paragraph{Sintering}
\label{sec:sintering}
Even at very lowest temperatures, the extremely small bond radius might allow diffusion processes like sintering to increase the contact area.
All contact conductivity models considered here assume that the contact quality ($r_{\mathrm{b}}/r_{\mathrm{g}}$) follows Hertz and JKR theory. However, sintering is expected to occur on icy moons \citep{Gundlach2018, Molaro2019, Mergny2024i}, which can significantly increase the contact area. Future models should couple estimated sintering geometries from multiphysics models such as LunaIcy \citep{Mergny2024i} with thermal conductivity models.
Such model could explain the higher thermal inertia of Jovian moons compared to those of the Saturnian moons, due to the overall warmer surface temperatures, as suggested by polarimetry of icy moons analogs \citep{Poch2018}.

At the same time, through Ostwald ripening, sintering in a non-uniform grain distribution would cause smaller grains to sublimate in favor of larger grains. As a result, a lower boundary for the grain radii would be added to the constraints in Figure~\ref{fig:all_TIs}, especially at higher temperatures ($>115$ K), where sintering is highly efficient. This would provide an even clearer picture of the allowed grain radii. \\

%Simple computation of the weigth $F=mg$ of the grains still shows that it is less than the van der Walls forces even at low temoperatures for small grains $<\SI{50}{um}$.

\section{Conclusion}

All icy moons in the solar system show remarkably low thermal inertia values, suggesting a regolith structure that deviates significantly from compact, homogeneous ice. 
In this study, we modeled contact and radiative thermal conductivity to identify the regions of parameter space where the derived thermal inertia remains below the observed low thermal inertia values (the $\Gamma < \Gamma_{\mathrm{LTIM}}$ relationship).  By systematically exploring these relationship for different icy moons conditions, we derived tight limits on the regolith’s porosity and grain radii relying on conservative assumptions.

Our analysis reveals that the uppermost layer of the regolith must have porosities exceeding 80\%, with a grain radius smaller than $\SI{1}{mm}$ for Saturnian moons and smaller than 500 $\mu m$ for Galilean icy moons. 
These findings are overall consistent with independent photometric \citep{Buratti1983,     Domingue1991, Buratti1995, Domingue1997, Hendrix2005}, spectroscopic \citep{Ligier2016, Mishra2021, King_2022, CruzMermy2023, CruzMermy2025}, and microwave observations \citep{Black2001a, Raza2026}, reinforcing the plausibility of our conclusions. 

Furthermore, we propose that the layered structure observed on Galilean moons can be explained by three key scenarios: deposition cover, degradation by impactors, and temperature gradient metamorphism. These mechanisms could collectively contribute to the observed thermal stratification of the regolith.

%Our constraints on regolith can be used by various method estimating parameters from the data, such as spectroscopy or photometry, by providing constraints on the grain size and porosity range to explore. It is also essential for lander missions, where understanding regolith porosity and grain size is crucial for instrument design and predicting thermal and mechanical behavior. 

While applied to icy moons, such work could also be used to better understand the extremely low thermal inertia of transneptunian bodies, like Pluto \citep{Bertrand2025}.

These constraints will be particularly valuable for photometry and spectroscopy studies relying on Bayesian inversion modeling of surface properties \citep{Belgacem2020, Mishra2021, CruzMermy2023, CruzMermy2025}, as they require sampling large parameter spaces. 
By putting constrains on microphysical properties, we narrow the free parameter space, which will enable more accurate inversion of compositional abundances and faster computation times. 
This has direct applications for future spectrometers MAJIS (Moons And Jupiter Imaging Spectrometer) onboard JUICE  \citep{Grasset2013} and MISE (Mapping Imaging Spectrometer for Europa) onboard Europa Clipper \citep{Pappalardo2024}, as well as for lander operations and landing site selection on highly porous icy surfaces (e.g., Voyager2050, ESA’s L4 mission).

\section{Appendix}

\subsection{Equivalence with the sintering adhesion stage}
\label{app:molarosint}

\citet{Ashby1974} introduced what they called ``\textit{Stage 0 sintering}'' as an instantaneous stage where an initial contact radius forms when two grains come into contact.
Following this formalism but applied to water ice, \citet{Molaro2019} uses the expression for the initial contact radius $r_{\mathrm{b}}^{S_0}$:
\begin{equation}
r_{\mathrm{b}}^{S_0} \approx \sqrt[3]{\dfrac{\gamma \,  r_{\mathrm{g}}^2}{10 \, G(T)} },
\end{equation}
where $G(T) = E/(2(1+\nu))$ is the ice shear modulus.
Comparing it with our Equation~\ref{eq:rb_express} for the initial contact radius based on \citet{Johnson1971} theory, we find that the expressions differ by a factor
\begin{equation}
\dfrac{r_{\mathrm{b}}^{S_0}}{r_{\mathrm{b}}} = \sqrt[3]{\dfrac{4(1+\nu)}{45 \pi}} \approx 0.335.
\end{equation}

Tracing the origin of  \citet{Ashby1974} Stage 0 expression, the authors refer to \citet{Easterling1972}, who used the same formula as our Equation~\ref{eq:rb_express}, derived from \citet{Johnson1971} theory. However, in their Equation 4, \citet{Easterling1972} introduced a simplification for metals, which \citet{Ashby1974} appears to have rounded, explaining the differences between $r_{\mathrm{b}}^{S_0}$ and $r_{\mathrm{b}}$.
Since both expressions derive from the same formalism in \citet{Johnson1971} theory, we recommend using Equation~\ref{eq:rb_express} for Stage 0 sintering calculations, as it avoids these approximations.

\subsection{Derivation to the general conductivity expression}
\label{app:cond_express}

\paragraph{Gundlach}
\citet{Gundlach2012} (in Eq. 12 of their manuscript) express the effective conductivity of ice as
\begin{equation}
    k = k_b(T) \left( \dfrac{9}{4} \dfrac{1-\nu^2}{E(T)} \pi \gamma(T) r^2 \right)^{1/3} \xi(r_{\mathrm{g}}, \phi)
\end{equation}
where we recognise the expression of $r_{\mathrm{b}}$ as given by equation \eqref{eq:rb_express}. Now using Eq. 18 of their paper, we obtain the formula
\begin{equation}
    k = k_b(T) \left[ f_1 \exp{(f_2 (1-\phi))} \right] \dfrac{r_{\mathrm{b}}}{r_{\mathrm{g}}}
\end{equation}

\paragraph{Snowpack and Adams}
Snowpack and \citet{Adams1993} express the effective conductivity as a porous media, when vapor heat transfer is negligible by
\begin{equation}
    k = \dfrac{n_{ca}}{n_{cl}} \dfrac{\pi^2}{32}r_{\mathrm{b}} k_b N_{\mathrm{c}}(\phi)
\end{equation}
where $n_{cl} = 1 / L_p $ with $L_p$ the mean pore length given by 
\begin{equation}
    L_p = \sqrt[3]{\dfrac{4 \pi r_{\mathrm{g}}^3}{3(1-\phi)}}.
\end{equation}
Since $n_{ca} = n_{cl}^2$ through simplification we find that the conductivity reduces to
\begin{equation}
    k = k_b \left[ \sqrt[3]{\dfrac{3}{4\pi}(1-\phi)} \dfrac{\pi^2}{32} N_{\mathrm{c}}(\phi) \right] \dfrac{r_{\mathrm{b}}}{r_{\mathrm{g}}}
\end{equation}
where we recognised the general expression of equation \eqref{eq:cond_sol_por} with the term in bracket being $S(\phi)$ as shown in the snowpack row of Table \ref{tab:sphis}.

\paragraph{Shoshany}
The author mentions that the addition of an Hertz factor should be included as an additional correction and then take account of the porous structure correction found in their paper. 
Hence we decided to include the Hertz factor to their $S(\phi)$ formula.

\paragraph{Coordination number}

\begin{figure}[htbp]
    \centering
    \includegraphics{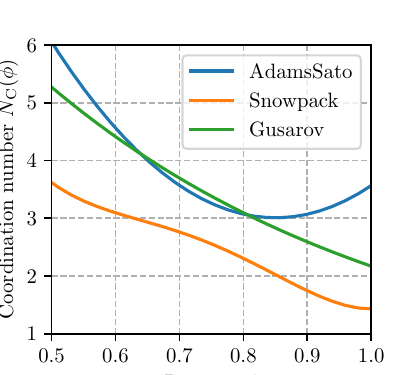}
    \caption{Coordination number from different studies in the literature. Their empirical formula are given in Table \ref{tab:sphis}.}
    \label{fig:coordnbr}
\end{figure}

Adams\&Sato, Gusarov and SNOWPACK models have different formulations of the coordination number as shown in Figure~\ref{fig:coordnbr}.
\label{app:coord}

\subsection{Alternative radiative conduction model}
\label{sec:BB_model}

\begin{figure}[htbp]
    \centering
    \includegraphics{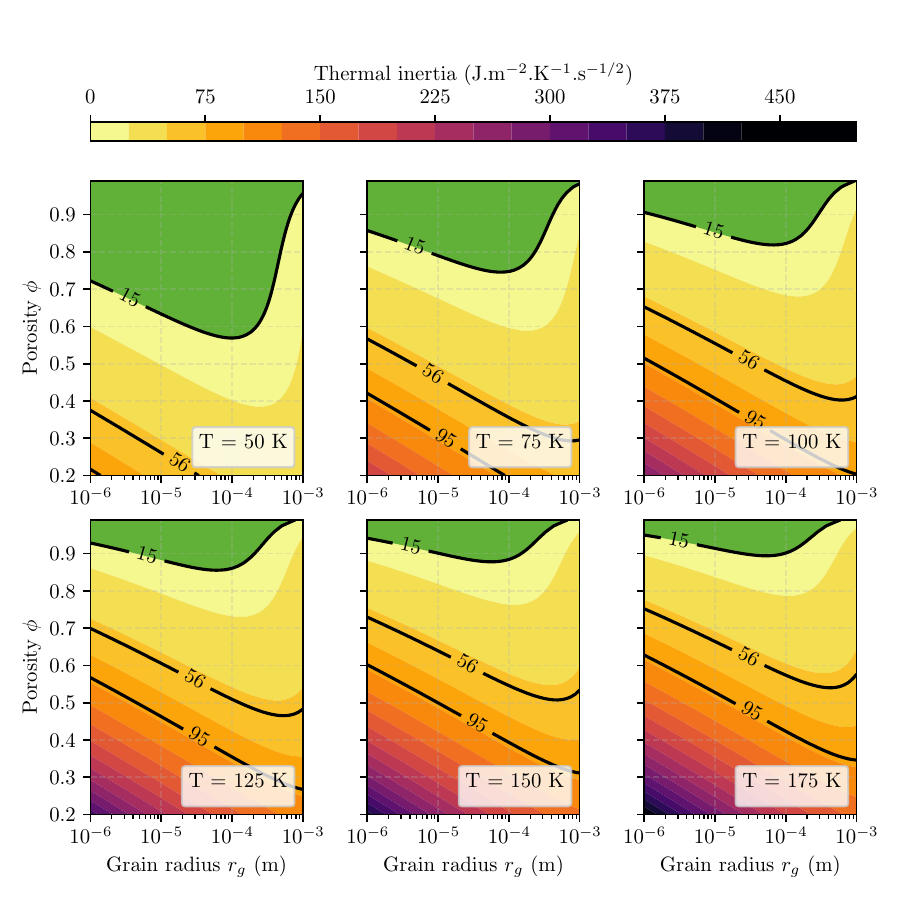}
    \caption{Equivalent to Figure~\ref{fig:param_allowed} but with \citet{Breitbach1980} radiative conductivity model.
    }
    \label{fig:param_allowed_BB}
\end{figure}

Alternative radiative conduction models have been proposed and have been compared by \citet{Ferrari2016}. 
To explore their impact on our results, we replaced Equation~\ref{eq:cond_rad} with the radiative model from \citet{Breitbach1980} (Equation 2 in \citet{Ferrari2016}).
As shown in Figure~\ref{fig:param_allowed_BB}, the overall porosity results remain consistent, but the \citet{Breitbach1980} model allows for larger grain radii, though still always below $\SI{1}{mm}$ for temperatures$ > 75 K$. For conservative studies, this figure may be used instead, as it provides a higher upper limit for grain radii.

\newpage
\section*{Open Research Section}
The models used here are described analytically in the manuscript and following the methods our results can be easily reproduced.

\section*{Conflict of Interest disclosure}
The authors declare there are no conflicts of interest for this manuscript.

\section*{Acknowledgments}
LL's research was supported by an appointment to the NASA Postdoctoral Program  administered by Oak Ridge Associated Universities  at the Jet Propulsion Laboratory,  California Institute of Technology, under a contract with the National Aeronautics and Space Administration (80NM0018D0004).

\bibliography{library}
\bibliographystyle{aasjournal}

\end{document}